\documentclass[10pt,a4paper]{article}

\usepackage[T1]{fontenc} 

\usepackage[english]{babel} 

\usepackage{abstract} 

\usepackage{titling} 

\usepackage{hyperref} 
\usepackage[square,numbers]{natbib}

\usepackage{graphicx}
\usepackage{siunitx}

\title{\LARGE \bf Switchbacks in the near-Sun magnetic field: \\ long memory and impact on the turbulence cascade}

\author{Thierry Dudok de Wit$^1$, Vladimir V. Krasnoselskikh$^1$, Stuart D. Bale$^{2,3,4}$, \\
John W. Bonnell$^2$, Trevor A. Bowen$^2$, Christopher H. K. Chen$^5$, Clara Froment$^1$, \\
Keith Goetz$^6$, Peter R. Harvey$^2$, Vamsee Krishna Jagarlamudi$^1$, Andrea Larosa$^1$, \\
Robert J. MacDowall$^7$, David M. Malaspina$^8$,  William H. Matthaeus$^9$, Marc Pulupa$^2$, \\
Marco Velli$^{10}$ and Phyllis L. Whittlesey$^2$
\\[1ex] 
\small $^1$ LPC2E, CNRS and University of Orl\'eans, 3A avenue de la Recherche Scientifique, Orl\'eans, France \\ 
\small $^2$ Physics Department, University of California, Berkeley, CA 94720-7300, USA \\ 
\small $^3$ Space Sciences Laboratory, University of California, Berkeley, CA 94720-7450, USA\\
\small $^4$ The Blackett Laboratory, Imperial College London, London, SW7 2AZ, UK\\
\small $^5$ School of Physics and Astronomy, Queen Mary University of London, London E1 4NS, UK\\
\small $^6$ School of Physics and Astronomy, University of Minnesota, Minneapolis, MN 55455, USA\\
\small $^7$ Solar System Exploration Division, NASA/Goddard Space Flight Center, Greenbelt, MD, 20771, USA\\
\small $^8$ Laboratory for Atmospheric and Space Physics, University of Colorado, Boulder, CO 80303, USA\\
\small $^9$ Department of Physics and Astronomy, University of Delaware, Newark, DE 19716, USA\\
\small $^{10}$  Institute of Geophysics \& University of California, Los Angeles, CA  90095-1567, USA\\
}

\renewcommand{\maketitlehookd}{%
\begin{abstract}
\noindent One of the most striking observations made by Parker Solar Probe during its first solar encounter is the omnipresence of rapid polarity reversals in a magnetic field that is otherwise mostly radial. These so-called switchbacks strongly affect the dynamics of the magnetic field. We concentrate here on their macroscopic properties. First, we find that these structures are self-similar, and have neither a characteristic magnitude, nor a characteristic duration. Their waiting time statistics shows evidence for aggregation. The associated long memory resides in their occurrence rate, and is not inherent to the background fluctuations. Interestingly, the spectral properties of inertial range turbulence differ inside and outside of switchback structures; in the latter the $1/f$ range extends to higher frequencies. These results suggest that outside of these structures we are in the presence of lower amplitude fluctuations with a shorter turbulent inertial range. We conjecture that these correspond to a pristine solar wind.

\end{abstract}
}

\date{\normalsize This article has been accepted for publication in:  \\
The Astrophysical Journal Supplement Series (2020), \\
\url{http://doi.org/10.3847/1538-4365/ab5853}} 

\begin{document}

\title{Switchbacks in the near-Sun magnetic field: \\ long memory and impact on the turbulence cascade}

\maketitle

\section{Introduction} 
\label{sec:intro}

In November 2018 Parker Solar Probe \citep{fox15} became the closest spacecraft ever to the Sun, reaching a distance of 35.7 $R_{\odot}$ or 0.166 AU from our star. This was the first of a series of 26 encounters during which the spacecraft will perform \textit{in situ} and remote sensing observations of the solar corona and gradually get as close as 9.8 $R_{\odot}$. This first encounter, however, has already provided a wealth of new results \citep{bale19,kasper19,howard19,mccomas19}. Among these is the omnipresence of swift reversals of the magnetic field, which is otherwise mostly radial. The elusive origin of these transient events, often called switchbacks, and their potential role in heating the solar wind has spurred active discussions. A global understanding of these switchbacks is gradually emerging as detailed comparisons of the different observations is progressing. Here, we consider these events from a more macroscopic point of view, and investigate what new insight can be gained from the dynamics of the magnetic field. 

\begin{figure}[!htb] 
    \begin{center}
    \includegraphics[width=0.95\textwidth]{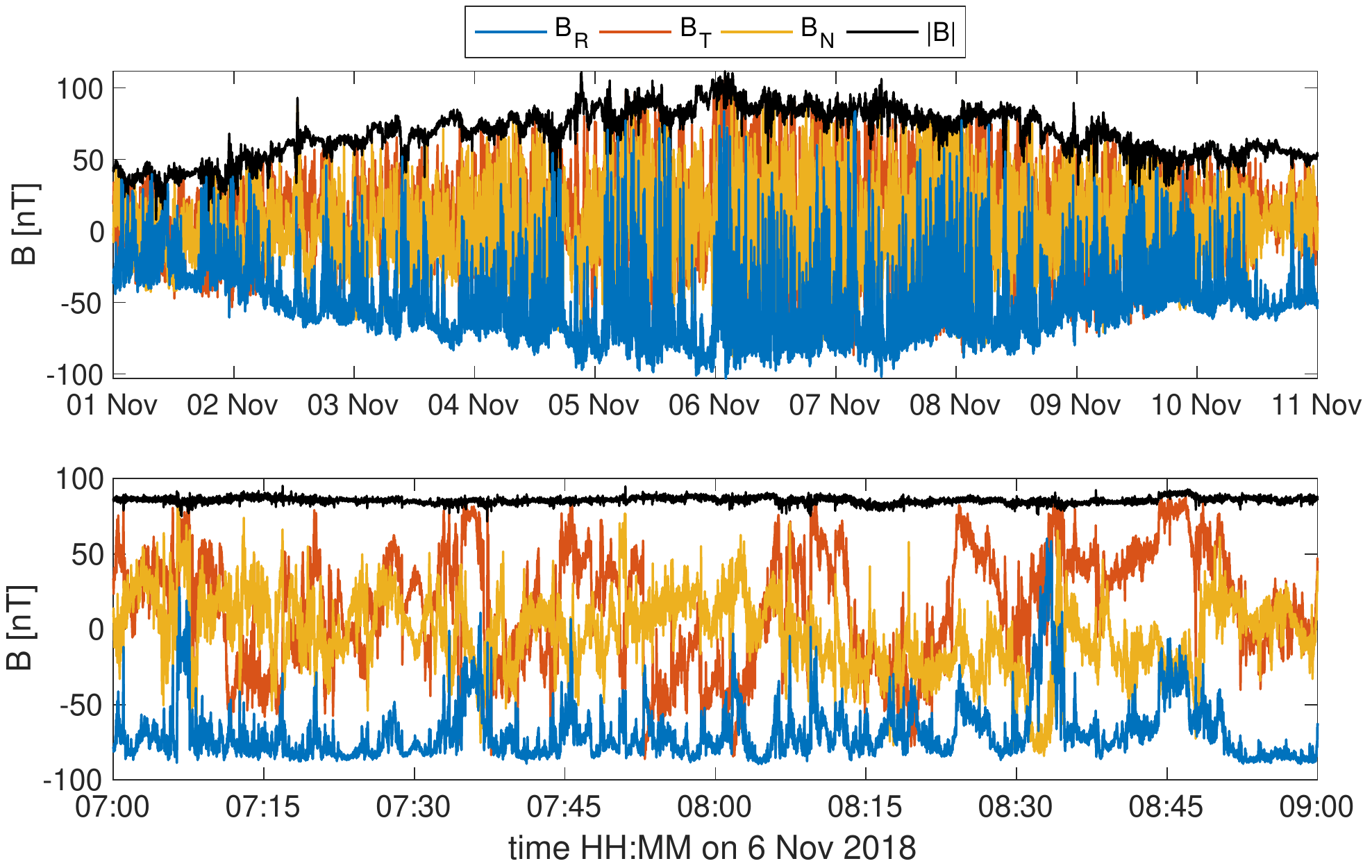}
    \end{center}
    
    \caption{Magnetic field for the whole interval of interest (upper plot) and for a two-hour excerpt (lower plot). An RTN coordinate system is used, see Sec.~\ref{sec:data}. \label{fig:rawdata}}
\end{figure}

The solar wind is a fascinating laboratory for studying waves and turbulence in collisionless plasmas \citep{bruno13}. However, by the time the wind that emanates from the Sun has reached the Earth's orbit (where most \textit{in situ} measurements are performed) it has already evolved dynamically for several days so that the properties of the pristine wind that Parker Solar Probe aims at observing are blurred. During its first perihelion pass, the spacecraft was almost co-rotating with the Sun and mostly observed a slow solar wind emerging from a small coronal hole located near the equator \citep{bale19,badman19}. For more than ten days around the first perihelion, the magnetic field was essentially radial with clear sunward polarity. This regularity was continuously interrupted by switchbacks, see Fig.~\ref{fig:rawdata}.

Switchbacks are characterized by their high alfv\'enicity, the absence of notable temperature changes and an increase in the wind velocity by approximately 20\% \citep{bale19,kasper19}. Likewise, the proton density also tends to increase inside these switchbacks. The  alfv\'{e}nicicity is attested by the low relative variability of the total magnetic field. The duration of these switchbacks ranges from less than a second to more than one hour. A remarkable result is the occasional presence between them of periods of quiet solar wind with low levels of magnetic field fluctuations and coherent wave activity \citep{bowen19,malaspina19}. 

Sudden reversals of the magnetic field have been observed for several decades. Early examples can be spotted in observations from the HELIOS mission  \citep{behannon81}. These reversals have since been studied at various distances from the Sun \citep{kahler96,ballegooien98, balogh99,yamauchi04,landi06,neugebauer12,neugebauer13,matteini14,borovsky16,horbury18}, and in simulations \citep{velli11}. 
However, the observations made by Parker Solar Probe stand out by 1) The absence of major velocity increase during switchbacks as compared to what has been observed, for example, by HELIOS at 0.3 AU \citep{borovsky16}; 2) Their observation in a slow (but alfv\'{e}nic) stream  and 3) The sharpness and omnipresence of these events, as if the coronal plasma was continuously transitioning between two metastable states: one in which the magnetic field is pointing sunward along the Parker spiral, and one in which it is deflected away from the spiral, sometimes by as much as \ang{180}. 

Several (and mutually not exclusive) explanations have been proposed for these switchbacks such as plasma jets associated with reconnection events deep inside the corona, or the crossing of kinked magnetic flux tubes, of closed magnetic loops, or of large-amplitude, non-compressive Alfv\'{e}n waves similar to those observed by \citet{belcher71}. The properties of the monodirectional energetic electron beams \citep{whittlesey19} and the inversion of the magnetic helicity \citep{mcmanus19} inside of the switchbacks give support to the idea that these structures are localized twists in the magnetic field and not polarity reversals or closed loops.

A consequence of the elusive origin of these switchbacks is the lack of consensus on their terminology. These structures are also called foldings, (intrasector) field reversals, jets, spikes, or deflections. They are truly deflections but we shall refer to them as switchbacks.

Interestingly, the presence of abrupt flow reversals is not unique to plasmas and has also been observed in hydrodynamic turbulence at high Rayleigh numbers \citep{araujo05} and in the geomagnetic field \citep{sorriso-valvo07b}. There are also analogies with dynamical systems that exhibit transitions between different states \citep{nicolis93,benzi05}. Metastable systems that are in contact with a (possibly external) perturbation have received considerable attention in the framework of non-equilibrium physics because they are capable of exhibiting effects that cannot be observed when the system is in thermal equilibrium \citep{gammaitoni98}. One of these effects is fluctuation amplification through stochastic resonance.

All these systems raise the same question: How can such transitions occur when their lifetime is so much longer than the characteristic timescales of the system? In the context of the solar wind, a related question is: Are these deflections part of the turbulent wavefield or are they signatures of perturbations that are generated by some external mechanism such as the moving photospheric footpoint of magnetic flux tubes? Beforehand, we should define what a switchback is. To answer these questions, we investigate here the dynamical properties of the magnetic field, following an exploratory approach.

This macroscopic exploration of the solar magnetic field is organized as follows: after presenting the data in Sec.~\ref{sec:data} we address their main properties in Sec.~\ref{sec:properties} followed by their waiting time statistics in Sec.~\ref{sec:waiting} and a focus on signatures of long memory in Sec.~\ref{sec:memory}. Finally we discuss the implications on our perception of solar wind turbulence in Sec.~\ref{sec:turbulence} and conclude in Sec.~\ref{sec:conclusions}.

\section{Data}
\label{sec:data}

The vector magnetic field is measured onboard Parker Solar Probe by the MAG magnetometer from the FIELDS consortium \citep{bale16}. At perihelion, the cadence of MAG goes up to 292.97 samples per second. Since we are interested in much lower frequencies, in what follows all observations will be decimated to 73.24 Hz.

Our time interval of interest is centered on the first perihelion pass of November 6, 2018 and runs from November 1st to November 10, 2018. During that 10-day interval the spacecraft was almost corotating with the Sun, so that temporal variations are primarily associated with spatial structures that are advected past the spacecraft with a typical solar wind speed of \SIrange{220}{450}{km/s} \citep{kasper19}. This  means that the crossing of thin structures such as filaments may take considerably longer than what other spacecraft usually see in the solar wind.

During that whole interval the photospheric footpoint of Parker Solar Probe was mostly located in the same equatorial coronal hole \citep{badman19}, causing the spacecraft to be immersed in a relative stable and highly alfv\'{e}nic slow solar wind. Outside of the interval the conditions progressively started changing with, for example, the occurrence of heliospheric perturbations such as a flux rope on November 12. The distance to the Sun varied from 51.4 to 35.7 $R_{\odot}$ and the median deviation of the magnetic field from the radial direction was \ang{12.4}. 

In the following we express the magnetic field in the RTN coordinate system: $R$ is radial and points away from the Sun, the tangential $T$ component is the cross product of the solar rotation vector with $R$; the normal $N$ component completes the right-handed set and points in the same direction as the solar rotation vector.

\section{Macroscopic properties of the switchbacks}
\label{sec:properties}

Figure~\ref{fig:rawdata} illustrates the main properties of magnetic field with conspicuous reversals in the radial $B_R$ component that are the hallmark of switchbacks. These deflections come in all magnitudes: some cover a few degrees only whereas the largest ones are fully anti-sunward. The high alfv\'{e}nicity is attested by the nearly constant total magnetic field. 

\begin{figure}[!htb] 
    \begin{center}
    \includegraphics[width=0.7\textwidth]{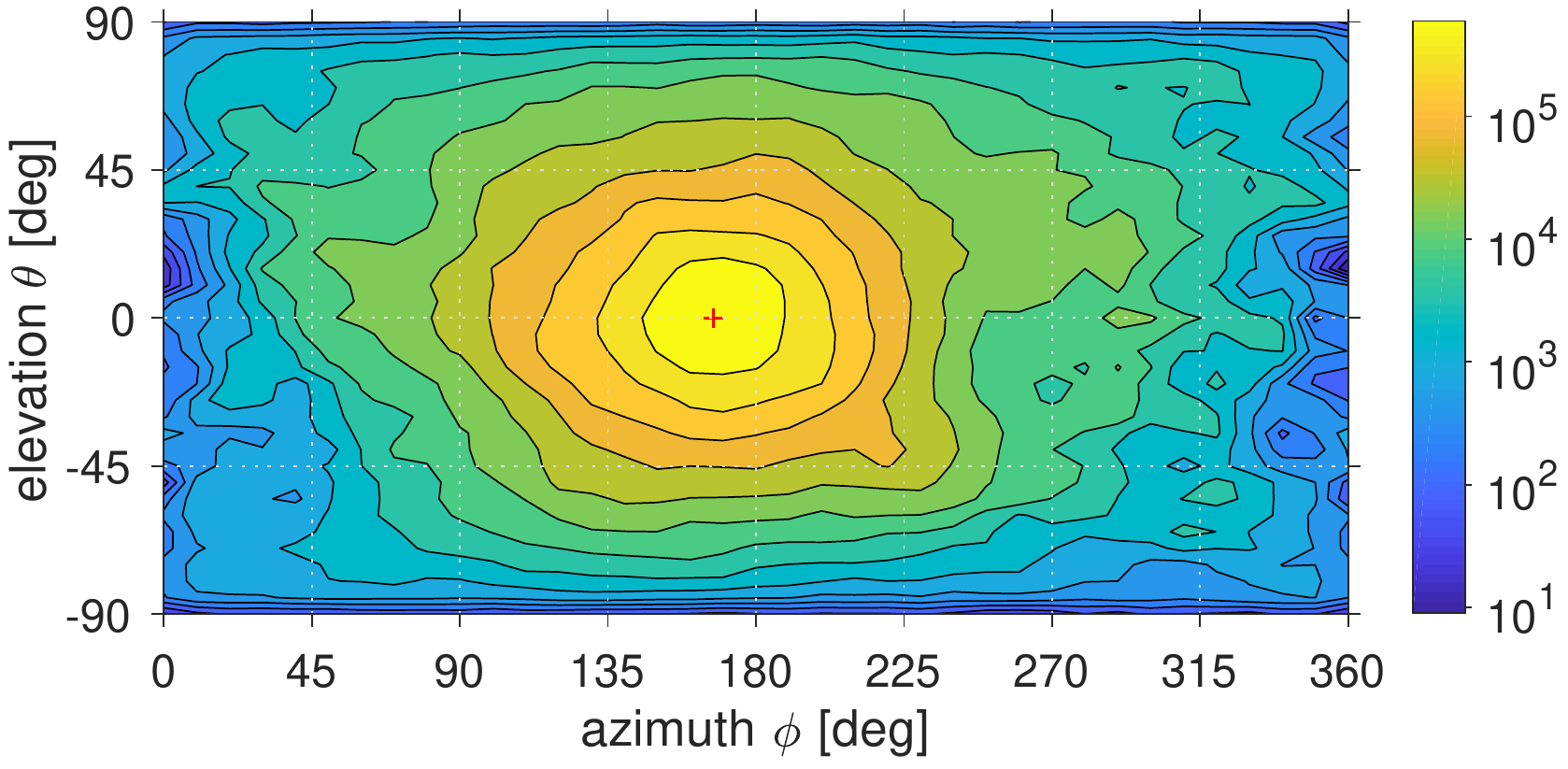}
    \end{center}
    
    \caption{2D histogram of the angular deflection of the magnetic field, for the full time interval. The orientation of the Parker spiral for prevailing solar wind conditions during the encounter is indicated by a cross. An azimuth of 180$^{\circ}$ corresponds to a sunward pointing of the magnetic field. \label{fig:isotropic}}
\end{figure}

Let us first determine whether these deflections have a preferred angular orientation. Figure~\ref{fig:isotropic} shows an occurrence histogram versus azimuth and elevation of all the 67'108'864 observations of the magnetic field that were recorded during the 10-day interval. What stands out is the isotropic distribution of the deflections with a strong concentration around an azimuth of \ang{170}; the latter coincides with the average position of the Parker spiral, which is indicated by a cross on the plot. We conclude that the bulk of the deflections has an isotropic distribution around the direction of the Parker spiral.

Figure~\ref{fig:isotropic} shows that large deflections are more sporadic than small ones. The largest ones also have a preferential tangential orientation \citep{kasper19,horbury19}. In what follows we shall concentrate on the salient features of the deflections and therefore ignore the weak anisotropy that affects the few largest events. Let us stress that this 2D distribution of the deflections does not significantly change when we estimate it separately for each of the ten days of the interval of interest. 

Thanks to the isotropy of the distribution of deflections we can now describe the dynamics of the magnetic field by means of one single parameter, which is its angular deflection with respect to the Parker spiral or mean magnetic field $\langle \vec{B} \rangle$. By analogy with magnetic lattices such as the Ising model \citep{peierls36}, let us consider the potential energy associated with each deflection
\begin{equation}
E_p = - \vec{B} \cdot \langle \vec{B} \rangle
\label{eq:potential_energy}
\end{equation}
where  $\langle \vec{B} \rangle$ is the prevalent magnetic field. Because of the high alfv\'{e}nicity, the magnitude of the magnetic field may be considered as being close to constant, mainly varying with the distance from the Sun. Therefore, the potential energy is mainly proportional to $- \cos \alpha$, where $\alpha$ is the angle of deflection. Based on this we introduce the dimensionless \textit{normalized deflection} $0 \le z \le 1$ as
\begin{equation}
 z = \frac{1}{2}(1 - \cos \alpha) \ ,
\end{equation}
as a convenient proxy for the deflection. Values that are close to 0 correspond to a ``ground state'' with a magnetic field that is aligned along the Parker spiral and pointing sunward. \citet{gosling09} had already noted that alfv\'{e}nic perturbations in the solar wind should be considered as deviations relative to a base value rather than to a time average. Conversely, $z$ reaches 1 when the magnetic field is pointing anti-sunward and along the Parker spiral. With this definition we can now decouple geometrical variations from weak but still omnipresent amplitude changes in the magnetic field.

\begin{figure}[!htb] 
    \begin{center}
    \includegraphics[width=0.95\textwidth]{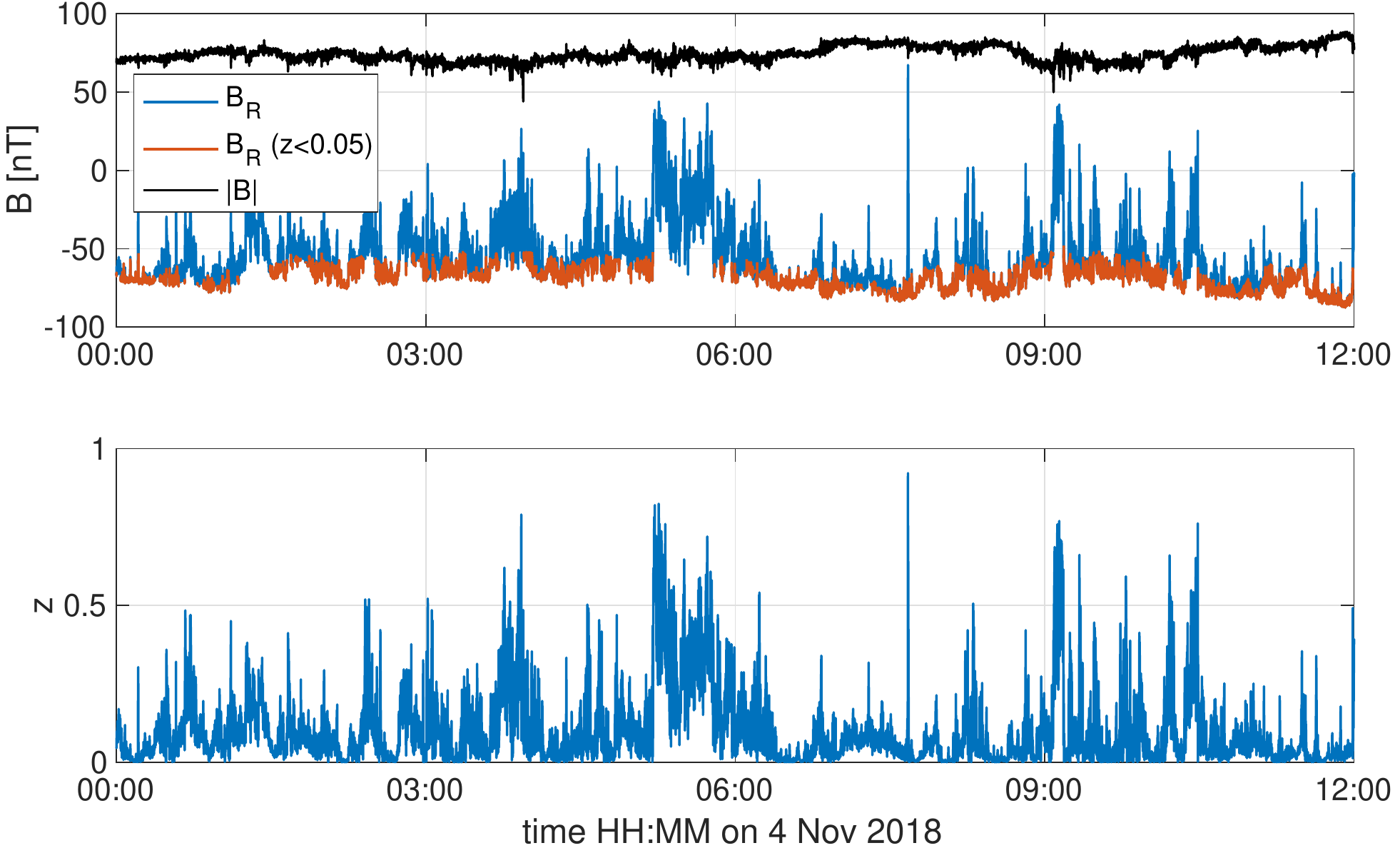}
    \end{center}
    
    \caption{Excerpt of the radial component of the magnetic field and the total magnetic field (upper plot) and the corresponding values of the normalized deflection $z$ (lower plot). Values of $B_R$ that correspond to a small normalized deflection of $z<0.05$ are highlighted. \label{fig:zdefinition}}
\end{figure}

Figure~\ref{fig:zdefinition} illustrates the difference between the radial magnetic field and the normalized deflection. Later in this study we shall discriminate deflections based on their value of $z$. 

Note that the prevalent orientation of the mean magnetic field fluctuates in time regardless of the presence of switchbacks. There are multiple reasons for this, including changes in  the velocity of the solar wind, and of the distance from the Sun. We find that 95\% of the azimuthal orientations of the prevalent magnetic field (estimated from a running median over 6-hour intervals) are comprised between \ang{-5.5} and \ang{19.1}. Likewise, 95\% of the elevations are between \ang{-10.6} and \ang{10.6}. These fluctuations cannot be neglected as they affect the value of the normalized deflection, see Eq.~\ref{eq:potential_energy}. Therefore, all deflections will be determined with respect to the direction of the prevalent magnetic field as estimated from a running median over 6-hour intervals. Taking longer intervals does not affect the conclusions of our study. Shorter intervals are not allowed since they must exceed the duration of the longest switchbacks. The jitter that results in the orientation of the prevalent field prevents us from distinguishing small variations in $z$. Therefore, in what follows, we shall systematically consider $z = 0.05$ as the threshold value below which the magnetic field is in a so-called ground state with small deflections.

The question now arises whether the distribution of $z$ shows any evidence for thresholds that would allow to define what a switchback is. This would not only help clarify the terminology but also give deeper insight into the physics. A bimodal distribution of $z$, for example, would indicate that the magnetic field oscillates between two metastable states. 

\begin{figure}[!htb] 
    \begin{center}
    \includegraphics[width=0.65\textwidth]{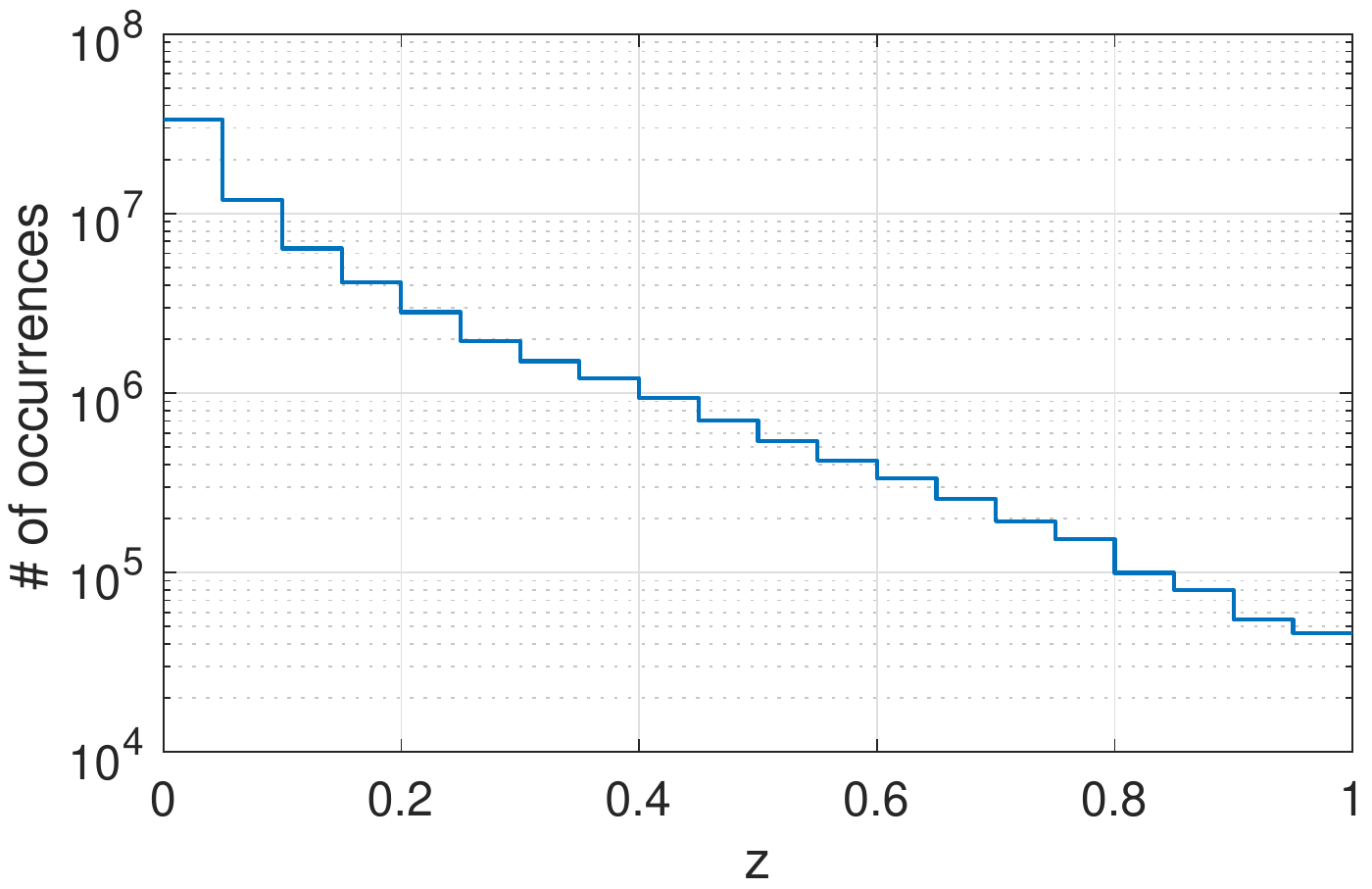}
    \end{center}
    
    \caption{Histogram of the values of $z$ for the 10-day time interval. \label{fig:pdfz}}
\end{figure}

The distribution of $z$ is shown in Fig.~\ref{fig:pdfz}, which reveals a monotonic and rather featureless distribution apart from a small excess of small deflections. We are therefore in the presence of a continuum of self-affine deflections with no particular threshold that would allow us to define quantitatively what a switchback is. We would have reached the same conclusions using the deflection angle $\alpha$. 
%

We made various attempts to identify switchbacks by other means, such as their spectral signature. \citet{farge15} have shown how coherent structures such as vortices can be separated from background fluctuations by means of the wavelet transform. We were unable to achieve a comparable separation between large switchbacks and the less deflected magnetic field, presumably because the spectral content does not provide enough leverage for disentangling the two.

Note that the distribution shown in Fig.~\ref{fig:pdfz} corresponds to deflections of all values of the magnetic field and not  individual switchback events. For the latter we need a definition  of what a switchback is. At the coarsest level one could define it as a deflection that exceeds the threshold value of $z=0.05$, and then drops again below it. With this definition we can investigate the distribution of the peak deflection per switchback. This distribution (not shown) is steeper than that of  Fig.~\ref{fig:pdfz}. However,  it is equally devoid of discontinuities and therefore corroborates the idea that there is no typical deflection for defining a switchback.

\section{Waiting time statistics}
\label{sec:waiting}

A key question is whether consecutive switchbacks tend to aggregate or are independent. For  this we consider the statistical distribution of waiting times (time elapsed between the end of a switchback and the onset of the next one) and of residence times (duration of a switchback) \citep{aschwanden14}. Numerous authors have investigated the waiting time distribution of intermittent events such as solar flares \citep[e.g.][]{wheatland02}, radio bursts \citep[e.g.][]{pulupa19} and discontinuities in the solar wind \citep [e.g.][]{greco09}. For a sequence of independent events that behave as a Poisson process, this distribution should be an exponential. In contrast, power law distributions arise when events are correlated. In practice, the distinction between the can easily be blurred by additional effects. Non-stationarity associated with time-varying flaring rates, for example, cause the distribution to be exponential for short durations and to roll over to a power law for long durations \citep{lepreti01,wheatland02}. 

\begin{figure}[!htb] 
    \begin{center}
    \includegraphics[width=0.75\textwidth]{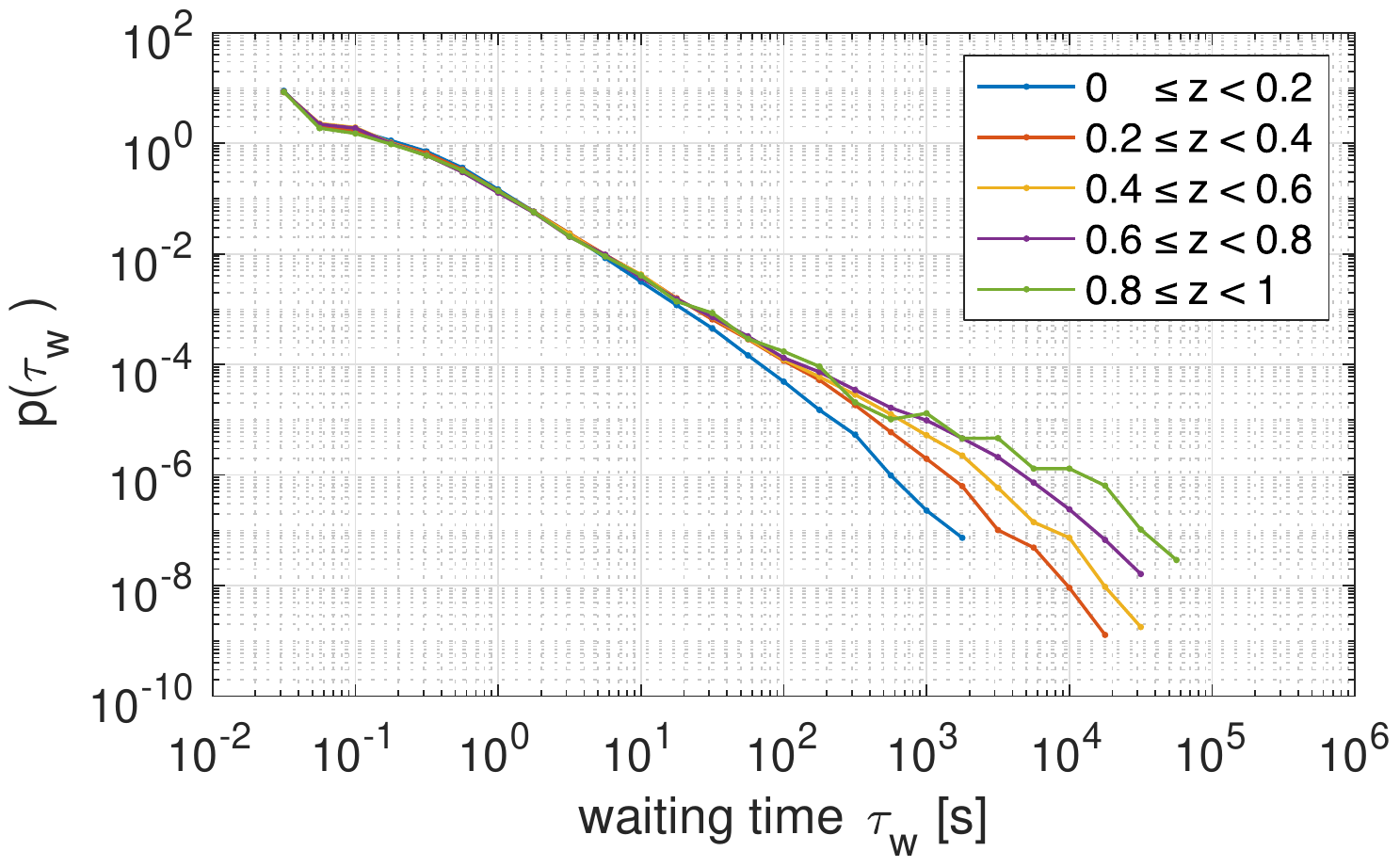}
    \end{center}
    
    \caption{Waiting time distribution of normalized deflection $z$ for different thresholds. \label{fig:pdftw}}

\end{figure}

Figure~\ref{fig:pdftw} summarizes the main results for the waiting time distribution using five thresholds of $z$. First, note how all the distributions collapse onto one single curve, except for the longest intervals that exceed hundreds of seconds. This is a strong indication that all deflections, whether small or large, have a common driver. Second, all distributions tend to follow a power law over several decades. The slope of this power law ranges between -1.4 and -1.6; similar values have been found in natural phenomena that exhibit clustering \citep{aschwanden14}. Note that the existence of such a power law implies that the mean waiting time is ill-defined.

Deviations from this power law for short waiting times can be explained by the presence of ion cyclotron waves \citep{bowen19} and interference noise from the spacecraft that distort the distribution. Interestingly, the power law breaks down too at long time scales. Although finite sample effects become increasingly important there is also a systematic depletion of long waiting times associated with small deflections. This may be due to a dual waiting time distribution with a mix of clustered deflections that are responsible for a power law and occasional quiescent intervals that exhibit smaller deflections. If the latter were randomly distributed, then the tail of the waiting time distribution should roll over to an exponential distribution, with fewer events as compared to what a power law should give \citep{aschwanden14}.

\begin{figure}[!htb] 
    \begin{center}
    \includegraphics[width=0.75\textwidth]{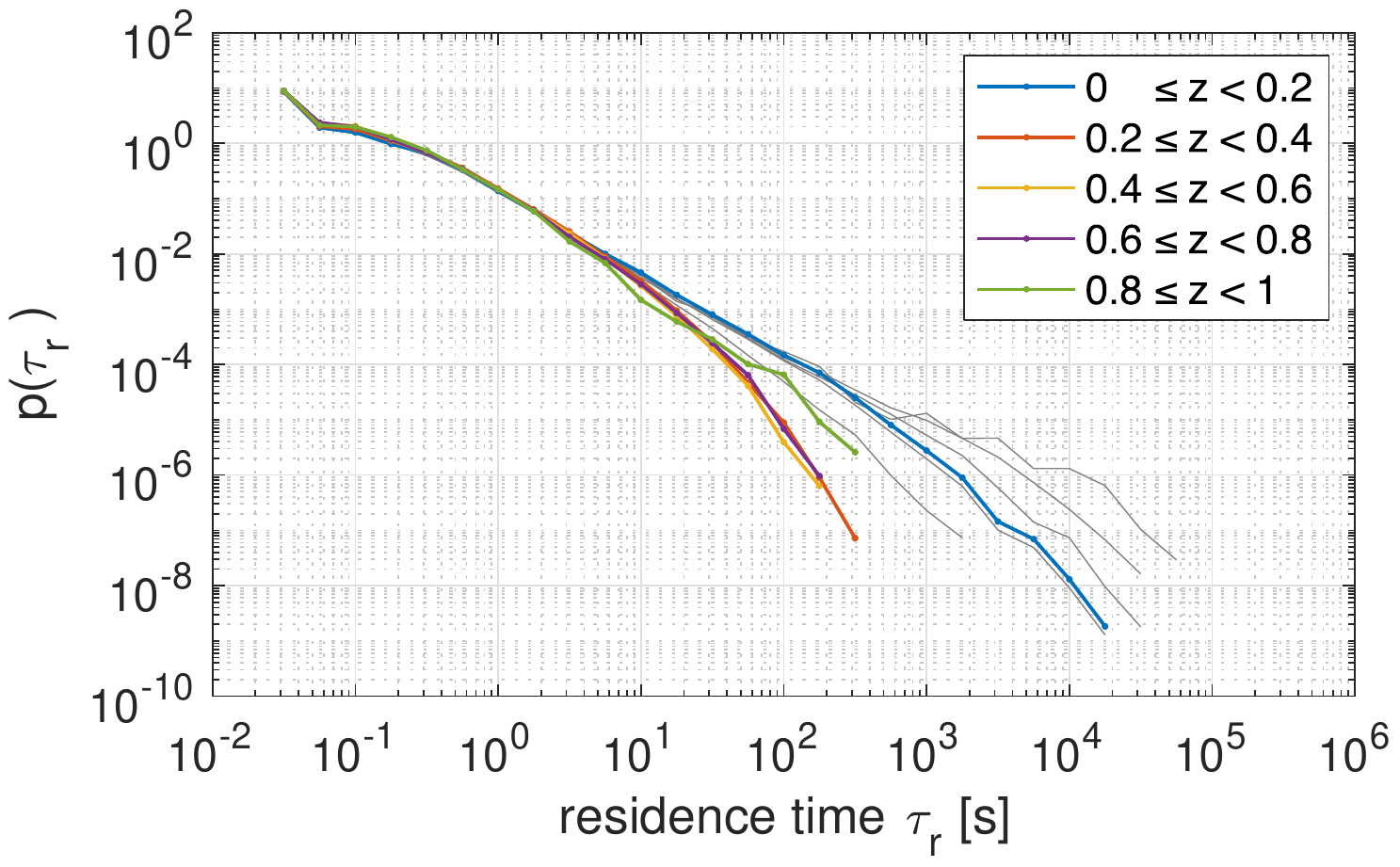}
    \end{center}
    
    \caption{Residence time distribution of of normalized deflection $z$ for different thresholds. The waiting time distributions from Fig.~\ref{fig:pdftw} are shown in gray, for comparison. \label{fig:pdftr}}\end{figure}

Further insight can be gained from the residence time distribution. Surprisingly, the waiting and residence time distributions are (to a first approximation) remarkably similar, see Fig.~\ref{fig:pdftr}. This important result suggests that the same physical processes govern the onset and the termination of a deflection. One would expect this for entangled filaments \citep{parker63} or flux tubes \citep{borovsky08} that are moving past the spacecraft. If, on the other hand, the deflections were due to outward propagating perturbations that are generated in the lower corona then the underlying instability would more likely give rise to different distributions. The same would occur for a plasma oscillating between two metastable states.

Figure~\ref{fig:pdftr} also reveals a conspicuous excess of long-duration intervals with small deflections, which indicates that there is a tendency for the magnetic field to remain in the ground state that has small deflections. Conversely, there is a relative deficit of large deflections.

A new picture emerges: in the previous Section we had concluded that there was a continuum of deflections. Here,  the waiting time distributions gives a more contrasted picture in which the ground state is relatively more frequent, as if the magnetic field was staying by default in this ground state, and leaving it only during deflections. The omnipresence of the deflections gives the false impression that there is no distinction between them and the ground state.
%

To summarize, to a first approximation, the magnetic field behaves as a monostable system with a preferential ground state. Deflections away and back to this ground state are ruled by the same physical processes and their omnipresence  leaves few opportunities for actually observing the ground state. The power law scaling of the waiting time distribution suggests that deflections aggregate, whatever their magnitude. There are no noticeable differences in these scalings when one-day intervals are chosen.

\section{Long memory effects}
\label{sec:memory}

The clustering of consecutive deflections suggests that there is long memory in the coronal magnetic field. This memory may have several origins. The most obvious one is a temporal correlation associated with the spatial coupling between adjacent magnetic flux tubes, similar to what is observed in spin lattices. These flux tubes, when mapped back into the photosphere, correspond to granules or supergranules whose small motion at the solar surface gets amplified because of the expansion of the flux tubes, and naturally leads to temporal correlations. Disentangling these different causes without multipoint observations is challenging.

To find clearer evidence for clustering between switchbacks we correlate each waiting time to its consecutive values. Since the waiting times vary by orders of magnitude, it is more appropriate to correlate their ranks, similar to the difference between Pearson and Spearman rank correlation functions \citep{press02}. Here, by estimating the autocorrelation function of the ranked values, we quantify whether waiting times of a given rank are likely to be followed values of a similar rank for the next event, the second next event, and so on.

\begin{figure}[!htb] 
    \begin{center}
    \includegraphics[width=0.65\textwidth]{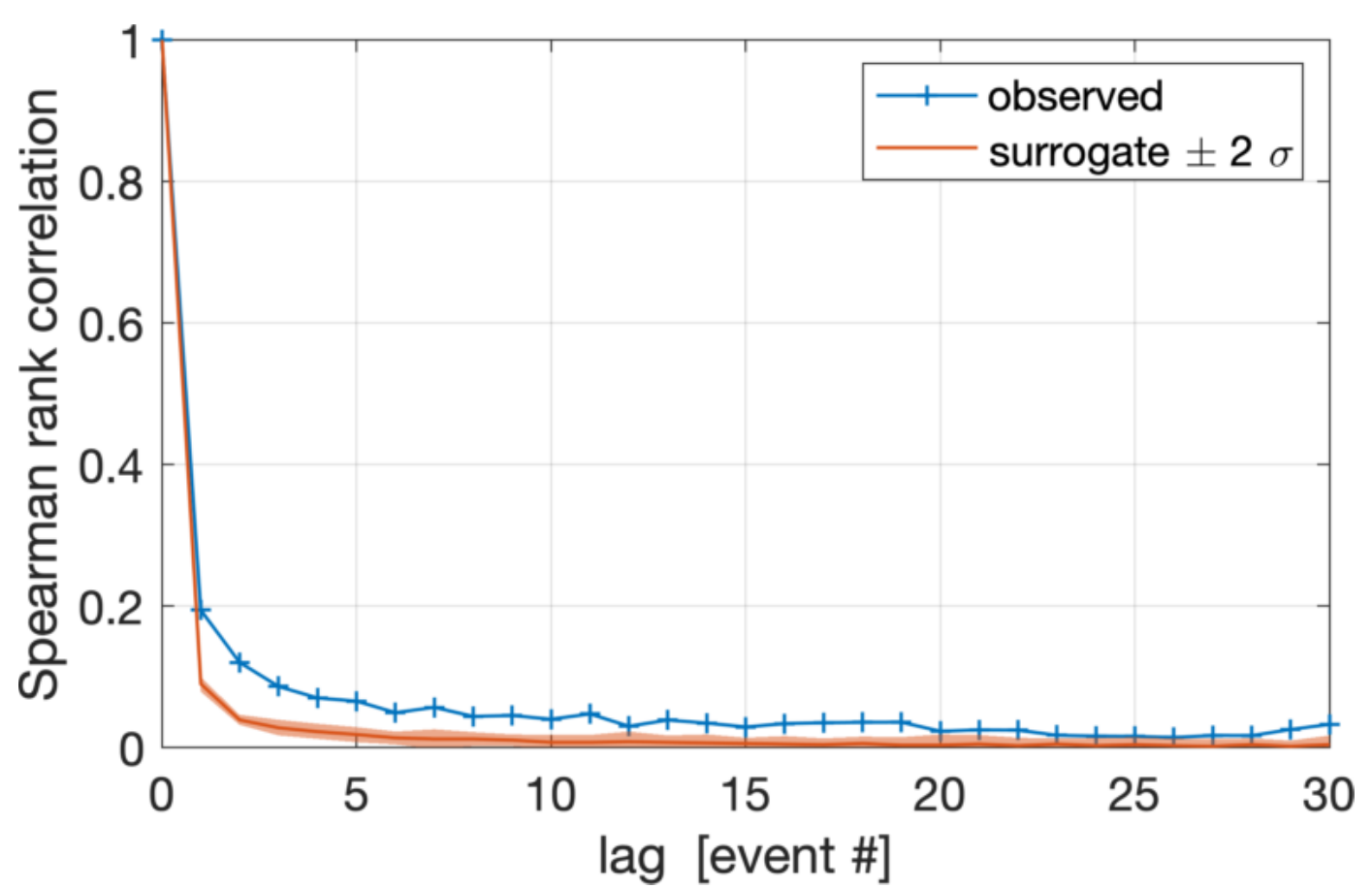}
    \end{center}
    
    \caption{Autocorrelation of the sequence of ranked waiting times (determined with a threshold value of $z=0.1$) for real observations and for surrogate data. For the latter, an average over 20 surrogates, with a $\pm 2\sigma$ interval are shown. Switchbacks that are shorter than 1 second have been excluded. \label{fig:autocorr}}
\end{figure}

Figure~\ref{fig:autocorr} illustrates the results of the autocorrelation of ranked waiting times for a threshold of $z=0.1$. Note the fast decay, which indicates that consecutive values rapidly differ from each other. The autocorrelation, however, does not drop to zero; its decay is well matched by a $ C(\textrm{lag}) \propto \textrm{lag}^{-1/2} $ scaling that is indicative of long-range correlations \citep{beran94}. 

The significance of this long tail of the autocorrelation function requires careful evaluation. Instead of considering confidence intervals we use a more stringent test and use surrogate data \citep{schreiber00} to answer the question: Do the nonzero correlations result from a nonlinear process that causes consecutive switchbacks to aggregate, or are they just a consequence of linear effects (i.e. the fact that nearby observations of the magnetic field are more likely to be similar)? More precisely, we test the null hypothesis that clustering in the normalized deflection $z$ is produced by a Gaussian linear stochastic process. To do so, we generate surrogate time series of $z$ that have the same probability distribution and same autocorrelation function (and therefore also same power spectral density) as the initial record $z$. The characteristics of these surrogates are remarkably similar to those of the original data (a trained eye can barely distinguish them) and yet, whatever nonlinear correlation may exist in the magnetic field has been destroyed in the surrogates. From these we estimate the waiting times whose autocorrelation is then compared to that of the original values.


The comparison of the two autocorrelation functions in Fig.~\ref{fig:autocorr} confirms the existence of a significant residual correlation between consecutive waiting times, which is absent in surrogate data. From this we conclude that there is a long memory in the sequence of switchbacks. Fig.~\ref{fig:autocorr} in addition shows that this memory persists at least for a dozen consecutive deflections.
%

These conclusions remain valid with other thresholds although the results are more contrasted when the threshold is between 0.05 and 0.2. To better diagnose the origin of this long memory let us now consider Detrended Fluctuation Analysis (DFA) \citep{beran94,kantelhardt01}, which is particularly suited for detecting long memory in non-stationary time series. 

In the following we consider $0$'th order DFA or DFA0 (also known as fluctuation analysis \citep{bryce12}) because we do not aim at removing trends. In DFA0 a sampled time series $x_t$ is sliced into $N_W$ non-overlapping windows of equal duration $\tau$ in each of which we calculate the variance. Finally we consider the root mean variance
\begin{equation}
 \sigma_x(\tau) = \sqrt{\frac{1}{N_W} \sum_{i=1}^{N_W} \big\langle \left(x_{t_i} - \langle x_t \rangle_{W_i} \right)^2 \big\rangle_{W_i}}
\end{equation}
in which $\langle \cdots \rangle_{W_i}$ stands for a time average over the interval $W_i: t_i \le t \le t_i + \tau$. For a self-affine process $\sigma(\tau) \propto \tau^{\mu}$, whose scaling exponent $\mu$ is related to the Hurst exponent. For uncorrelated fluctuations such as those resulting from Brownian motion $\mu = \frac{1}{2}$ whereas for serial correlation one should have $\mu > \frac{1}{2}$. Conversely $\mu < \frac{1}{2}$ implies anticorrelation.  These scaling exponents embody a close correspondence with physical scenarios \citep{metzler09}.

Using DFA we can now compare the magnetic field with the normalized deflection, and also select intervals according to the magnitude of the deflections. This enables us to determine whether the evidence for long memory is the consequence of the switchbacks only or if it is intrinsic to the complete sequence of observations. In addition, by applying DFA to $z$ and to the different components of the magnetic field, we should be able to determine whether the origin of long memory is geometrical and resides the orientation of the magnetic field or if variations of the magnetic field amplitude contribute to it. 

\begin{figure}[!htb] 
    \begin{center}
    \includegraphics[width=0.75\textwidth]{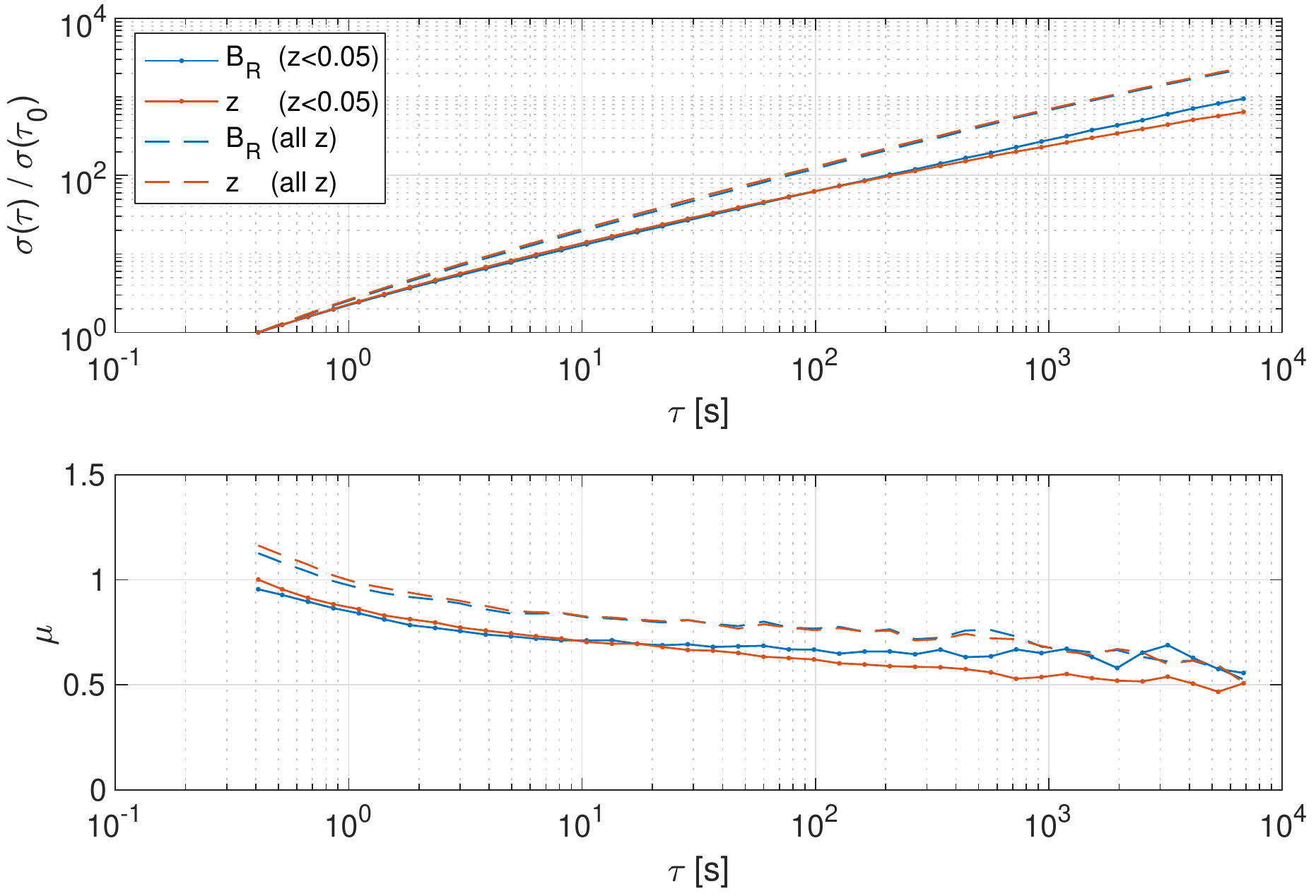}
    \end{center}
    
    \caption{Detrended fluctuation analysis (DFA0) applied to $B_R$ and to the normalized deflection $z$, for quiescent intervals and for active intervals. The upper plot shows $\sigma(\tau)$ normalized to its value for the shortest time; the lower plot shows the corresponding scaling exponent $\mu$. \label{fig:dfa}}

\end{figure}

The results of DFA analysis are summarized in Fig.~\ref{fig:dfa}. As before we focus on the radial component of the magnetic field and consider here two cases: quiescent conditions that should be close to the ground state ($z<0.05$) and active conditions ($0\le z\le 1$), both for $z$ and for $B_R$.

The lower plot of Fig.~\ref{fig:dfa} displays the main quantity of interest, which is scaling exponent $\mu$. For quiescent conditions $\mu$ remains close to 0.5, except at short time scales ($\tau < \SI{10}{\sec}$) where coherent wave activity is likely to affect the results. Such  a value of $\mu$ implies the absence of long memory in both $B_R$ or $z$ when the magnetic field is its ground state; it is also the signature of regular diffusion. 

If, however, we apply DFA to active conditions, then the scaling exponent is systematically larger and saturates around 0.6. Such values are the signature of long memory, or at least longer lasting memory than in the quiescent regime. Numerous studies have shown how anisotropic magnetic fields such as those associated with meandering field lines can lead to anomalous superdiffusive particle transport \citep{jokipii69,pommois01,perrone13}. In this context, the aggregation of the switchbacks also provides favorable conditions for non-diffusive transport.

Interestingly, the scaling exponents are nearly identical for $z$ and for $B_R$; consequently, long memory effects occur in the angular deflections rather than in the fluctuations of the magnetic field that come on top of these. In other words, the long memory resides in the way switchbacks occur as events while the background magnetic field fluctuations have limited memory.

The picture that emerges  reveals a magnetic field whose ground state has no long memory, in contrast to the intermittent deflections that tend to aggregate and therefore exhibit long memory. These distinct properties corroborate the idea that the two have different origins. This surprising emergence of long memory in transient events rather than in the ground state has received considerable attention in the context of climate science because of its implications on predictability \citep{koutsoyiannis03}. \citet{franzke15} have shown that it is precisely the rate of switching between different states (in our case, the occurrence of deflections) that can generate long memory and be responsible for non-stationarity in a system that is otherwise memoryless. Recently, \citet{sato19} have provided yet another explanation for this behavior by showing that particles that are sampled randomly either from a memoryless system with  Brownian motion (i.e. our quiescent ground state) or from a non-chaotic system in which particles aggregate (i.e. switchbacks) exhibit anomalous diffusion with scaling exponents $\mu>0.5$.

\section{Impact on turbulence cascade}
\label{sec:turbulence}

\begin{figure}[!htb] 
    \begin{center}
    \includegraphics[width=0.8\textwidth]{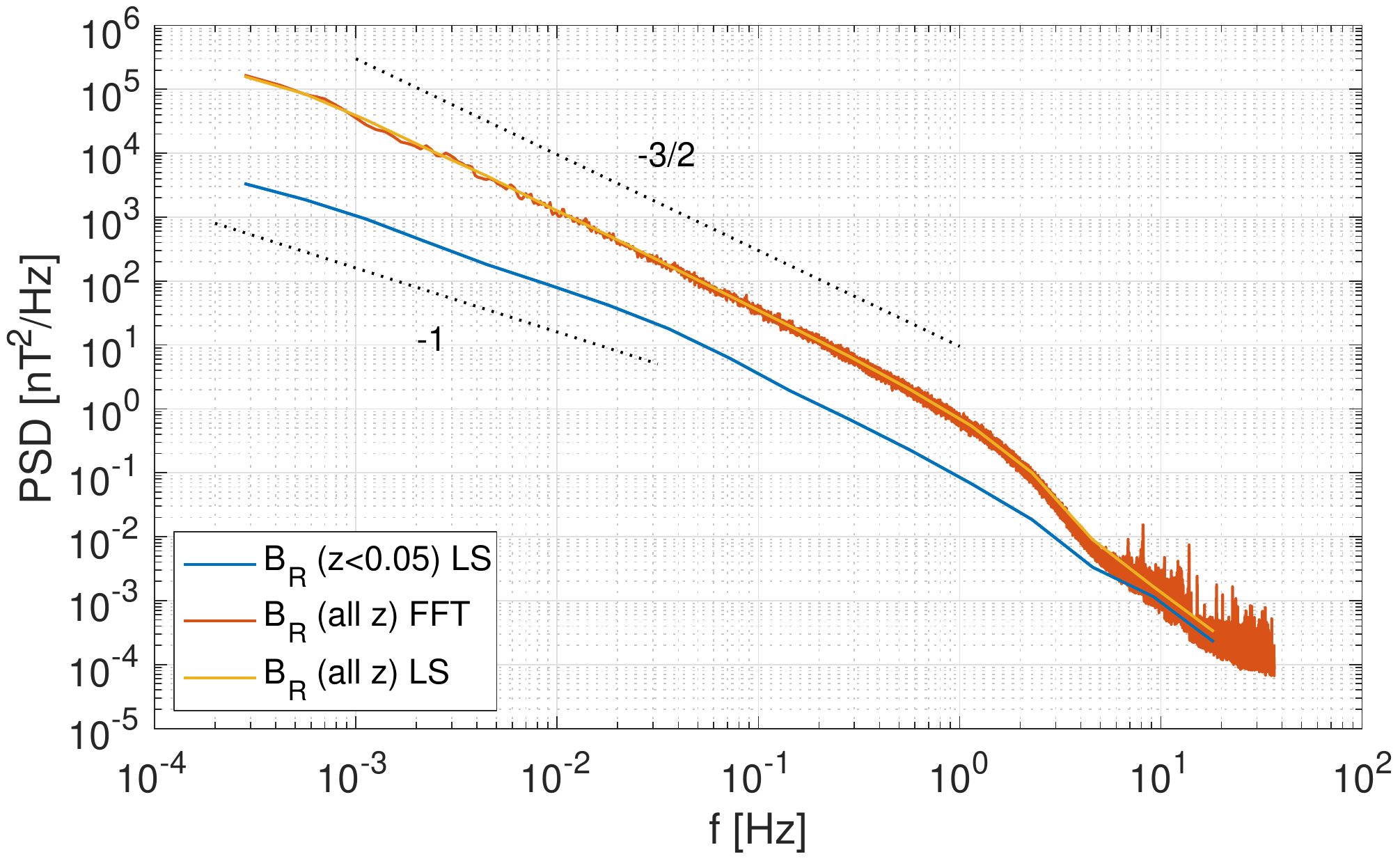}
    \end{center}
    
    \caption{Power spectral density of the radial component $B_R$ of the magnetic field for quiescent ($z<0.05$) and active (all $z$) regimes. For the latter, we overplot the results obtained by using the Fast Fourier Transform (FFT) and the Lomb-Scargle (LS) method to confirm their good agreement. \label{fig:spec1}}
\end{figure}

Now that there is evidence for the magnetic field fluctuations to have different origins depending on their deflection, let us  turn to their spectral properties and determine whether switchbacks can be considered as being part of the turbulent wavefield. For this, we consider the power spectral density of the different components of the magnetic field, and also of $z$, for specific thresholds on $z$. The results are summarized Fig.~\ref{fig:spec1} for quiescent conditions ($z<0.05$) and active ones ($0 \le z \le 1$). The precise value of the threshold has no major impact on our conclusions as long as it properly isolates the quiescent state.

Since the quiescent state occurs only sporadically we use the Lomb-Scargle method \citep{press02,vio13} to estimate the power spectral density from irregularly sampled data and obtain access to time scales that exceed the duration of individual quiescent intervals. We systematically process in parallel a benchmark sequence with a known spectrum in order to check the validity of our results.

Figure~\ref{fig:spec1} summarizes the main results by comparing the power spectral density of the radial component $B_R$ for quiescent and for active conditions. In both regimes we observe power laws $P(f) \propto f^{\beta}$ that extend over more than one decade. Frequencies above \SI{1}{Hz} are discarded because they belong to the kinetic range. In addition they are affected by narrow-band noise from the spacecraft reaction wheels. The properties below \SI{1}{Hz} in active conditions are typical for MHD turbulence, with an inertial range whose spectral index is close to $\beta = -3/2$, preceded by a so-called $1/f$ range whose slope is close to $\beta = -1$ \citep{chen19}. The break frequency between the two is located around \SI{0.001}{Hz}, which is consistent with the fluctuation level of the magnetic field \citep{matteini19}.

\begin{figure}[!htb] 
    \begin{center}
    \includegraphics[width=0.8\textwidth]{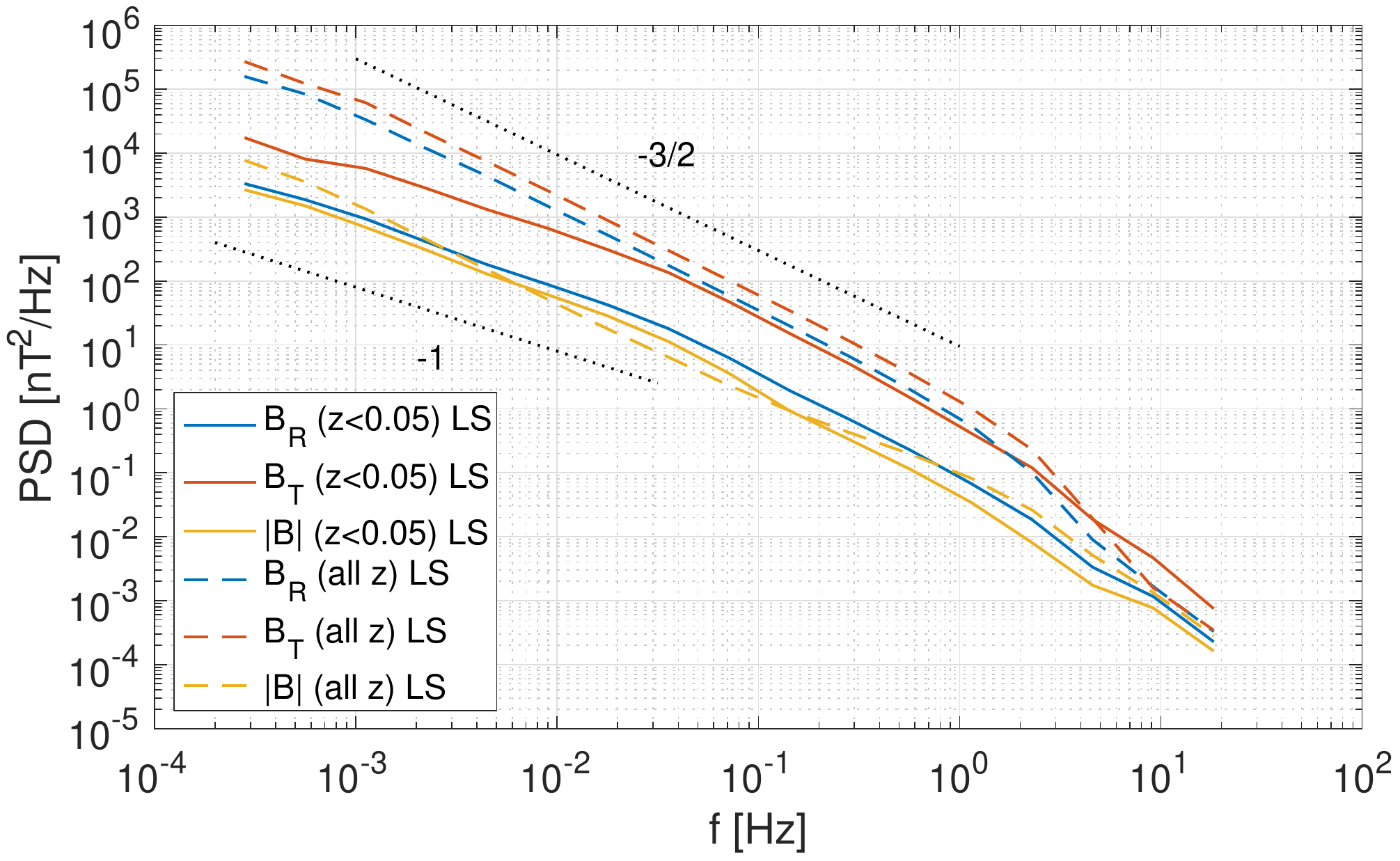}
    \end{center}
    
    \caption{Same as Fig.\ref{fig:spec1}, now comparing two components of the magnetic field and the total field for quiescent and active regimes. All estimates are based on the Lomb-Scargle (LS) method. \label{fig:spec2}}

\end{figure}

The active and quiescent regimes have comparable spectral indices between \SI{0.05}{Hz} and \SI{1}{Hz}, which approximately corresponds to the inertial range, see Table~\ref{tab:spectral_indices}). At lower frequencies, however, the spectral index changes because the break frequency moves up from \SI{0.001}{Hz} to \SI{0.05}{Hz}. The $1/f$ range is considered as an energy reservoir of the turbulent cascade and consists of a mix of non fully developed turbulence and coherent structures that still carry the signature of the sources of the wind \citep{bruno13,matteini18}. The 50-fold increase in the break frequency may then suggest that in the quiescent regime the turbulent cascade has had only time to develop a short inertial range. In other words, the quiescent regime carries a signature of a more pristine and un-evolved solar wind.

\begin{table}[!htb]
\caption{Spectral index of the power spectral density of $B_R$ estimated for active ($0 \le z \le 1$) and quiescent ($z<0.05$) conditions. }
\begin{center}
\begin{tabular}{ccc}
Frequency range [Hz] & Conditions & $\beta$ \\ \hline
$10^{-3}$ -- $3\cdot10^{-2}$ & Active & $-1.50 \pm 0.03$ \\
$10^{-3}$ -- $3\cdot10^{-2}$ & Quiescent & $-1.07 \pm 0.03$ \\
$5 \cdot 10^{-2}$ -- $1$ & Active & $-1.63 \pm 0.05$ \\
$5 \cdot 10^{-2}$ -- $1$ & Quiescent & $-1.70 \pm 0.12$ \\
\end{tabular}
\end{center}
\label{tab:spectral_indices}
\end{table}%

The same pattern arises in the power spectral density of the transverse components and the modulus of the magnetic field, see Fig.~\ref{fig:spec2}, with some subtle differences that are beyond the scope of this study. Other effects may explain such a shift in the break frequency. Since the fluctuation level is lower in the quiescent regime, the turbulent eddies take longer to interact, which naturally shifts the break frequency upwards. While this effect certainly plays a role, we note that by selecting larger thresholds of $z$ (and thus allowing for eddies of larger amplitude) the break frequency does not immediately drop, as would be expected.

If quiescent conditions indeed correspond to a pristine solar wind, then how deep in the corona are these fluctuations generated? To answer this question we consider the correlation length of the magnetic field along the radial direction. The autocorrelation function of $B_R$ decays monotonically and its $e$-folding time  provides a coarse estimate of the decay time. By using a solar wind velocity model \citep{zouganelis04} with the observed solar wind velocity as input, we can then approximately locate the source of the fluctuations. Let us stress that the decay time is indicative only because the magnetic field is non-stationary and so the autocorrelation function is biased \citep{jagarlamudi19}. 

In the active regime the correlation time is of the order of a few minutes only because because erratic deflections rapidly destroy the correlation. In the quiescent regime, however, the autocorrelation decays  more slowly with a characteristic decay time of $T=10 \pm 7$ [h]. The large uncertainty reflects the dispersion of the values as estimated from a sequence of 2-day intervals. Note that this decay time is biased and therefore should be considered as a lower bound.

The long decay time associated with the quiescent regime corresponds to a source located below 20 $R_{\odot}$. This is lower than the Alfv\'{e}n point, which is estimated to be located between 25 and 30 $R_{\odot}$ \citep{kasper19}. We conclude that the fluctuations that are seen during the quiescent regime correspond to the transit of highly elongated structures that most likely originate deep inside the corona. Possibly as deep as the photosphere. 

\section{Discussion and Conclusions}
\label{sec:conclusions}

Our results lead to an apparent contradiction. On one hand we observe a continuum of deflections in the magnetic field, which would suggest that there is one single regime only, with a seamless continuum between a quiescent state with small deflections, and large switchbacks. Here we ignore coherent wave activity that occurs on top of this and is further addressed in \citep{bowen19,malaspina19}. 

On the other hand, we find evidence for long memory in the deflections only and not in the quiescent regime. From this we conclude that the magnetic field is more likely to exhibit two regimes: a quiescent ground state with no long memory, and occasional perturbations (deflections) that do exhibit long memory.

The distinction between the two pictures brings us back to the initial question: Are switchbacks part of the turbulent wavefield or should they be considered as external perturbations? This distinction between the two, if it exists, is blurred by several effects such the small variations in the orientation of the Parker spiral.  

Several indicators support the idea that switchbacks and the quiescent ground state are distinct. First, the similarity between the waiting and residence time distributions suggests that switchbacks correspond to a deviation from the ground state, which excludes the existence of a bistable state as observed, for example, in some plasma instabilities \citep{passot06}. Second, the relative excess of occurrences of long quiescent regimes pleads in favor of their different nature. A third indicator is the different spectral density of $B_R$, which  we interpret as a signature of un-evolved solar wind for the quiescent regime.  

These indicators do not exclude alternate explanations such as one in which deflections would be remnants of large switchbacks that have evolved into fully developed turbulence. Here the turbulent wavefield conserves imprints of the initial drivers. This may explain one of our puzzling results, which is the similarity of the spectral indices as inferred from the active regime and as reported for  the inertial range of turbulence in the inner heliosphere \citep{chen19} or in corresponding MHD turbulence models \citep{chandran15}, as if a mix of switchback with pristine un-evolved solar wind equals inertial range turbulence. Note that this similarity could also be a mere coincidence.  Indeed, random transitions between states can easily give rise to power spectral densities that follow a power law with a spectral index located between -1 and -2. Incidentally, this highlights the ambiguity of the spectral index, for which completely different physical processes can give rise to comparable values. 

If these switchbacks are distinct from the quiescent ground state, do the two eventually merge into a fully developed turbulence energy cascade? The study by \citet{borovsky16} suggests that switchbacks do not evolve after 0.3 AU, as if they had stopped interacting with the surrounding turbulent wavefield. Several studies suggest that switchbacks observed closer to the Sun by Parker Solar Probe do interact with the ambient magnetic field:  \citet{krasnoselskikh19} argue that wave activity at the boundary of switchbacks may contribute to mixing and eventually to their weakening while \citet{tenerani19}, using simulations, conjecture that they originate in the lower corona and survive out to Parker Solar Probe distances. All these results support the idea that the concentration of switchbacks is higher at the perihelion of Parker Solar Probe than farther out. Note, however, that our perception is biased by the near-corotation of the spacecraft with the Sun, which favors the observation of radial structures advected over the spacecraft rather than the  crossing of magnetic flux tubes.

At this stage we are unable to nail down the precise origin of these switchbacks, for which a microscopic investigation of the wave and particle properties is required. However, our macroscopic analysis favors mechanisms in which these structures can be described as temporary departures from a ground state with a strong spatial correlation in the source of these deflections. Both switchbacks and the quiescent solar wind must originate deep in the corona. This picture is more difficult to reconcile with jets that would be produced by reconnection events lower in the corona because the rate of occurrence of the latter is unlikely to exhibit long memory. For the same reason, it is more difficult to reconcile with a model that would involve self-organized criticality. On the contrary, the picture is compatible with the meandering of kinked magnetic field lines.

To conclude, let us summarize the main facts, and then proceed with more speculative results.

\begin{itemize}
 \item First we find that magnetic switchbacks come in all durations (seconds to hours) and all angular deflections (\ang{0} to \ang{180}) with respect to the Parker spiral. They are best observed in radial component of the magnetic field, on which this study focuses. The bulk of the distribution (more than 98\%) is isotropically distributed around the Parker spiral and can therefore be conveniently described by one single dimensionless quantity, which we call the normalized deflection. From the distribution of deflections we conclude that there is no crisp definition of what a switchback is. We are rather in the presence of a continuum of deflections that could typically result from a fractal distribution of kinked magnetic flux tubes.
 
 \item Both the waiting time and residence time distribution of these deflections tend to follow a power law and are remarkably similar. This suggests that consecutive deflections tend to aggregate. Furthermore, the onset and termination of a deflection must be governed by the same physical processes. Finally, the magnetic field does not behave as a system with different metastable states, but rather as a monostable system that moves away from its ground state for each deflection. This  explains the relative excess of long quiescent time intervals (i.e. with small deflections of typically $z<0.1$). 
  
 \item Autocorrelation and detrended fluctuation analysis confirm the presence of long memory in the magnetic field. Interestingly, this long memory manifests itself in the occurrence of deflections and not in the fluctuations that are observed during the quiescent regime. Such anisotropies of the magnetic field with long-range correlations offer are known to generate anomalous particle transport \citep{pommois01}. 
 
Recently, \citet{sato19} have  shown how a similar behavior arises in systems that result from a mix between   memoryless Brownian motion and non-chaotic aggregation. The long memory we observe is most likely associated with the strong spatial connection between adjacent magnetic flux tubes and their common photospheric footpoints \citep[e.g.][]{borovsky08}.
 
 \item While the power spectral density of the magnetic field data is consistent with scaling laws found for  the inner heliosphere \citep{chen19,matteini19} the quiescent regime exhibits a longer $1/f$  range. We conjecture that this quiescent regime corresponds to a more pristine solar wind whose spectral energy cascade has only had time to develop a short inertial range. From its correlation length we estimate its source to be located below 20 $R_{\odot}$, i.e.  well below  the Alfv\'{e}n point. 
 
Surprisingly, the spectral properties of the mix between switchback and quiescent solar wind equals that of classical solar wind turbulence, which raises questions as to how much these switchbacks actually interact and merge with the  background turbulence. 
 
\end{itemize}


Taken together, these results suggest that switchbacks are probably not inherently part of the turbulent wavefield but are remnants of strong deflections that occurred deeper down in the corona and are connected by their common origin. Their spectral signature mimics that of inertial range turbulence, which suggest that they are gradually merging with it. In contrast, the spectral signature of the quiescent magnetic field  between switchbacks is reminiscent of a more pristine and less evolved solar wind. These properties, and in particular the long memory of the rate of occurrence of switchbacks are compatible with a physical picture in which these structures are the signature of kinked magnetic flux tubes that are moving past the spacecraft. The combined long memory and high anisotropy of these events offer favorable conditions for anomalous particle transport.

\subsection*{Acknowledgments}

The authors thank the FIELDS team of Parker Solar Probe as well as the Parker Solar Probe mission operations and spacecraft engineering teams at the Johns Hopkins University Applied Physics Laboratory for their support. The FIELDS experiment on the Parker Solar Probe spacecraft was designed and developed under NASA contract NNN06AA01C. All data used in this work are available on the FIELDS data archive: \url{http://fields.ssl.berkeley.edu/data/}. This research has made use of NASA’s Astrophysics Data System Bibliographic Services. TD, CF, VJ, VK and AL acknowledge funding from CNES. CHKC is supported by STFC Ernest Rutherford Fellowship ST/N003748/2. SDB acknowledges the support of the Leverhulme Trust Visiting Professorship program.





\begin{thebibliography}{63}
\providecommand{\natexlab}[1]{#1}
\providecommand{\url}[1]{\texttt{#1}}
\expandafter\ifx\csname urlstyle\endcsname\relax
  \providecommand{\doi}[1]{doi: #1}\else
  \providecommand{\doi}{doi: \begingroup \urlstyle{rm}\Url}\fi

\bibitem[{Araujo} et~al.(2005){Araujo}, {Grossmann}, and {Lohse}]{araujo05}
F.~F. {Araujo}, S.~{Grossmann}, and D.~{Lohse}.
\newblock {Wind Reversals in Turbulent Rayleigh-B{\'e}nard Convection}.
\newblock \emph{Physical Review Letters}, 95\penalty0 (8):\penalty0 084502,
  August 2005.
\newblock \doi{10.1103/PhysRevLett.95.084502}.

\bibitem[Aschwanden et~al.(2014)Aschwanden, Crosby, Dimitropoulou, Georgoulis,
  Hergarten, McAteer, Milovanov, Mineshige, Morales, Nishizuka, Pruessner,
  Sanchez, Sharma, Strugarek, and Uritsky]{aschwanden14}
Markus~J. Aschwanden, Norma~B. Crosby, Michaila Dimitropoulou, Manolis~K.
  Georgoulis, Stefan Hergarten, James McAteer, Alexander~V. Milovanov, Shin
  Mineshige, Laura Morales, Naoto Nishizuka, Gunnar Pruessner, Raul Sanchez,
  A.~Surja Sharma, Antoine Strugarek, and Vadim Uritsky.
\newblock 25 years of self-organized criticality: Solar and astrophysics.
\newblock \emph{Space Science Reviews}, 98\penalty0 (1--4):\penalty0 47--66,
  2014.
\newblock \doi{10.1007/s11214-014-0054-6}.

\bibitem[Badman et~al.(2020)Badman, Bale, Oliveros, Pansenco, Velli, Stansby,
  Buitrago-Casas, R{\'e}ville, Bonnell, Case, de~Wit, Goetz, Harvey, Kasper,
  Korreck, Larson, Livi, MacDowall, Malaspina, Pulupa, Stevens, , and
  Whittlesey]{badman19}
Samuel~T. Badman, Stuart~D. Bale, Juan C.~Martinez Oliveros, Olga Pansenco,
  Marco Velli, David Stansby, Juan~C. Buitrago-Casas, Victor R{\'e}ville,
  John~W. Bonnell, Anthony~W. Case, Thierry~Dudok de~Wit, Keith Goetz, Peter~R.
  Harvey, Justin~C. Kasper, Kelly~E. Korreck, Davin~E. Larson, Roberto Livi,
  Robert~J. MacDowall, David~M. Malaspina, Marc Pulupa, Michael~L. Stevens, ,
  and Phyllis~L. Whittlesey.
\newblock {Magnetic connectivity of the ecliptic plane within 0.5 AU : PFSS
  modeling of the first PSP encounter}.
\newblock \emph{The Astrophysical Journal Supplement Series}, in press, 2020.
\newblock \doi{10.3847/1538-4365/ab4da7}.

\bibitem[Bale et~al.(2016)Bale, Goetz, Harvey, Turin, Bonnell, Dudok~de Wit,
  Ergun, MacDowall, Pulupa, Andre, Bolton, Bougeret, Bowen, Burgess, Cattell,
  Chandran, Chaston, Chen, Choi, Connerney, Cranmer, Diaz-Aguado, Donakowski,
  Drake, Farrell, Fergeau, Fermin, Fischer, Fox, Glaser, Goldstein, Gordon,
  Hanson, Harris, Hayes, Hinze, Hollweg, Horbury, Howard, Hoxie, Jannet,
  Karlsson, Kasper, Kellogg, Kien, Klimchuk, Krasnoselskikh, Krucker, Lynch,
  Maksimovic, Malaspina, Marker, Martin, Martinez-Oliveros, McCauley, McComas,
  McDonald, Meyer-Vernet, Moncuquet, Monson, Mozer, Murphy, Odom, Oliverson,
  Olson, Parker, Pankow, Phan, Quataert, Quinn, Ruplin, Salem, Seitz, Sheppard,
  Siy, Stevens, Summers, Szabo, Timofeeva, Vaivads, Velli, Yehle, Werthimer,
  and Wygant]{bale16}
S.~D. Bale, K.~Goetz, P.~R. Harvey, P.~Turin, J.~W. Bonnell, T.~Dudok~de Wit,
  R.~E. Ergun, R.~J. MacDowall, M.~Pulupa, M.~Andre, M.~Bolton, J.-L. Bougeret,
  T.~A. Bowen, D.~Burgess, C.~A. Cattell, B.~D.~G. Chandran, C.~C. Chaston,
  C.~H.~K. Chen, M.~K. Choi, J.~E. Connerney, S.~Cranmer, M.~Diaz-Aguado,
  W.~Donakowski, J.~F. Drake, W.~M. Farrell, P.~Fergeau, J.~Fermin, J.~Fischer,
  N.~Fox, D.~Glaser, M.~Goldstein, D.~Gordon, E.~Hanson, S.~E. Harris, L.~M.
  Hayes, J.~J. Hinze, J.~V. Hollweg, T.~S. Horbury, R.~A. Howard, V.~Hoxie,
  G.~Jannet, M.~Karlsson, J.~C. Kasper, P.~J. Kellogg, M.~Kien, J.~A. Klimchuk,
  V.~V. Krasnoselskikh, S.~Krucker, J.~J. Lynch, M.~Maksimovic, D.~M.
  Malaspina, S.~Marker, P.~Martin, J.~Martinez-Oliveros, J.~McCauley, D.~J.
  McComas, T.~McDonald, N.~Meyer-Vernet, M.~Moncuquet, S.~J. Monson, F.~S.
  Mozer, S.~D. Murphy, J.~Odom, R.~Oliverson, J.~Olson, E.~N. Parker,
  D.~Pankow, T.~Phan, E.~Quataert, T.~Quinn, S.~W. Ruplin, C.~Salem, D.~Seitz,
  D.~A. Sheppard, A.~Siy, K.~Stevens, D.~Summers, A.~Szabo, M.~Timofeeva,
  A.~Vaivads, M.~Velli, A.~Yehle, D.~Werthimer, and J.~R. Wygant.
\newblock {The FIELDS Instrument Suite for Solar Probe Plus}.
\newblock \emph{Space Science Reviews}, 204\penalty0 (49):\penalty0 1--34,
  2016.
\newblock ISSN 1572-9672.
\newblock \doi{10.1007/s11214-016-0244-5}.
\newblock URL \url{http://dx.doi.org/10.1007/s11214-016-0244-5}.

\bibitem[Bale et~al.(2019)Bale, Badman, Bonnell, Bowen, Burgess, Case, Cattell,
  Chandran, Chaston, Chen, Drake, de~Wit, Eastwood, Ergun, Farrell, Fong,
  Goetz, Goldstein, Goodrich, Harvey, Horbury, Howes, Kasper, Kellogg,
  Klimchuk, Korreck, Krasnoselskikh, Krucker, Laker, Larson, MacDowall,
  Maksimovic, Malaspina, Martinez-Oliveros, McComas, Meyer-Vernet, Moncuquet,
  Mozer, Phan, Pulupa, Raouafi, Salem, Stansby, Stevens, Szabo, Velli, Woolley,
  and Wygant]{bale19}
S.~D. Bale, S.~T. Badman, J.~W. Bonnell, T.~A. Bowen, D.~Burgess, A.~W. Case,
  C.~A. Cattell, B.~D.~G. Chandran, C.~C. Chaston, C.~H.~K. Chen, J.~F. Drake,
  T.~Dudok de~Wit, J.~P. Eastwood, R.~E. Ergun, W.~M. Farrell, C.~Fong,
  K.~Goetz, M.~Goldstein, K.~A. Goodrich, P.~R. Harvey, T.~S. Horbury, G.~G.
  Howes, J.~C. Kasper, P.~J. Kellogg, J.~A. Klimchuk, K.~E. Korreck, V.~V.
  Krasnoselskikh, S.~Krucker, R.~Laker, D.~E. Larson, R.~J. MacDowall,
  M.~Maksimovic, D.~M. Malaspina, J.~Martinez-Oliveros, D.~J. McComas,
  N.~Meyer-Vernet, M.~Moncuquet, F.~S. Mozer, T.~D. Phan, M.~Pulupa, N.~E.
  Raouafi, C.~Salem, D.~Stansby, M.~Stevens, A.~Szabo, M.~Velli, T.~Woolley,
  and J.~R. Wygant.
\newblock {Highly structured slow solar wind emerging from an equatorial
  coronal hole}.
\newblock \emph{Nature}, 2019.
\newblock \doi{10.1038/s41586-019-1818-7}.
\newblock URL \url{https://doi.org/10.1038/s41586-019-1818-7}.

\bibitem[{Balogh} et~al.(1999){Balogh}, {Forsyth}, {Lucek}, {Horbury}, and
  {Smith}]{balogh99}
A.~{Balogh}, R.~J. {Forsyth}, E.~A. {Lucek}, T.~S. {Horbury}, and E.~J.
  {Smith}.
\newblock {Heliospheric magnetic field polarity inversions at high heliographic
  latitudes}.
\newblock \emph{Geophysical Research Letters}, 26:\penalty0 631--634, 1999.
\newblock \doi{10.1029/1999GL900061}.

\bibitem[{Behannon} and {Burlaga}(1981)]{behannon81}
K.~W. {Behannon} and L.~F. {Burlaga}.
\newblock {Alfven Waves and Alfvenic Fluctuations in the Solar Wind}.
\newblock In H.~{Rosenbauer}, editor, \emph{Solar Wind 4}, pages 374--385,
  Garching, 1981. Max-Planck-Institute f{\"u}r Aeronomie.

\bibitem[{Belcher} and {Davis}(1971)]{belcher71}
J.~W. {Belcher} and L.~{Davis}, Jr.
\newblock {Large-amplitude Alfv{\'e}n waves in the interplanetary medium, 2}.
\newblock \emph{Journal of Geophysical Research}, 76:\penalty0 3534, 1971.
\newblock \doi{10.1029/JA076i016p03534}.

\bibitem[{Benzi}(2005)]{benzi05}
R.~{Benzi}.
\newblock {Flow Reversal in a Simple Dynamical Model of Turbulence}.
\newblock \emph{Physical Review Letters}, 95\penalty0 (2):\penalty0 024502,
  July 2005.
\newblock \doi{10.1103/PhysRevLett.95.024502}.

\bibitem[Beran(1994)]{beran94}
Jan Beran.
\newblock \emph{{Statistics for long-memory processes}}.
\newblock {Chapman and Hall}, New York, 1994.

\bibitem[{Borovsky}(2008)]{borovsky08}
J.~E. {Borovsky}.
\newblock {Flux tube texture of the solar wind: Strands of the magnetic carpet
  at 1 AU?}
\newblock \emph{Journal of Geophysical Research (Space Physics)}, 113:\penalty0
  A08110, August 2008.
\newblock \doi{10.1029/2007JA012684}.

\bibitem[{Borovsky}(2016)]{borovsky16}
J.~E. {Borovsky}.
\newblock {The plasma structure of coronal hole solar wind: Origins and
  evolution}.
\newblock \emph{Journal of Geophysical Research (Space Physics)}, 121:\penalty0
  5055--5087, June 2016.
\newblock \doi{10.1002/2016JA022686}.

\bibitem[{Bowen} and et~al.(2020)]{bowen19}
T.~{Bowen} and et~al.
\newblock {Inner Heliosphere Observations of Ion Scale Electromagnetic Waves}.
\newblock \emph{The Astrophysical Journal Supplement Series}, submitted, this
  volume, 2020.

\bibitem[{Bruno} and {Carbone}(2013)]{bruno13}
R.~{Bruno} and V.~{Carbone}.
\newblock {The Solar Wind as a Turbulence Laboratory}.
\newblock \emph{Living Reviews in Solar Physics}, 2:\penalty0 4, September
  2013.
\newblock \doi{10.12942/lrsp-2013-2}.

\bibitem[{Bryce} and {Sprague}(2012)]{bryce12}
R.~M. {Bryce} and K.~B. {Sprague}.
\newblock {Revisiting detrended fluctuation analysis}.
\newblock \emph{Scientific Reports}, 2:\penalty0 315, March 2012.
\newblock \doi{10.1038/srep00315}.

\bibitem[{Chandran} et~al.(2015){Chandran}, {Schekochihin}, and
  {Mallet}]{chandran15}
B.~D.~G. {Chandran}, A.~A. {Schekochihin}, and A.~{Mallet}.
\newblock {Intermittency and Alignment in Strong RMHD Turbulence}.
\newblock \emph{The Astrophysical Journal}, 807:\penalty0 39, July 2015.
\newblock \doi{10.1088/0004-637X/807/1/39}.

\bibitem[Chen et~al.(2020)Chen, Bale, Bonnell, Borovikov, Bowen, Burgess, Case,
  Chandran, {Dudok de Wit}, Goetz, Harvey, Kasper, Klein, Korreck, Larson,
  Livi, MacDowall, Malaspina, Mallet, McManus, Moncuquet, Pulupa, Stevens, and
  Whittlesey]{chen19}
C.~H.~K. Chen, S.~D. Bale, J.~W. Bonnell, D.~Borovikov, T.~A. Bowen,
  D.~Burgess, A.~W. Case, B.~D.~G. Chandran, T.~{Dudok de Wit}, K.~Goetz, P.~R.
  Harvey, J.~C. Kasper, K.~G. Klein, K.~E. Korreck, D.~Larson, R.~Livi, R.~J.
  MacDowall, D.~M. Malaspina, A.~Mallet, M.~D. McManus, M.~Moncuquet,
  M.~Pulupa, M.~Stevens, and P.~Whittlesey.
\newblock {The Evolution and Role of Solar Wind Turbulence in the Inner
  Heliosphere}.
\newblock \emph{The Astrophysical Journal Supplement Series}, submitted, this
  volume, 2020.

\bibitem[{Farge} and {Schneider}(2015)]{farge15}
M.~{Farge} and K.~{Schneider}.
\newblock {Wavelet transforms and their applications to MHD and plasma
  turbulence: a review}.
\newblock \emph{Journal of Plasma Physics}, 81\penalty0 (6):\penalty0
  435810602, December 2015.
\newblock \doi{10.1017/S0022377815001075}.

\bibitem[Fox et~al.(2015)Fox, Velli, Bale, Decker, Driesman, Howard, Kasper,
  Kinnison, Kusterer, Lario, Lockwood, McComas, Raouafi, and Szabo]{fox15}
N.~J. Fox, M.~C. Velli, S.~D. Bale, R.~Decker, A.~Driesman, R.~A. Howard, J.~C.
  Kasper, J.~Kinnison, M.~Kusterer, D.~Lario, M.~K. Lockwood, D.~J. McComas,
  N.~E. Raouafi, and A.~Szabo.
\newblock The solar probe plus mission: Humanity's first visit to our star.
\newblock \emph{Space Science Reviews}, 204\penalty0 (1--4):\penalty0 7--48,
  November 2015.
\newblock ISSN 1572-9672.
\newblock \doi{10.1007/s11214-015-0211-6}.
\newblock URL \url{http://dx.doi.org/10.1007/s11214-015-0211-6}.

\bibitem[{Franzke} et~al.(2015){Franzke}, {Osprey}, {Davini}, and
  {Watkins}]{franzke15}
C.~L.~E. {Franzke}, S.~M. {Osprey}, P.~{Davini}, and N.~W. {Watkins}.
\newblock {A Dynamical Systems Explanation of the Hurst Effect and Atmospheric
  Low-Frequency Variability}.
\newblock \emph{Scientific Reports}, 5:\penalty0 9068, March 2015.
\newblock \doi{10.1038/srep09068}.

\bibitem[{Gammaitoni} et~al.(1998){Gammaitoni}, {H{\"a}nggi}, {Jung}, and
  {Marchesoni}]{gammaitoni98}
L.~{Gammaitoni}, P.~{H{\"a}nggi}, P.~{Jung}, and F.~{Marchesoni}.
\newblock {Stochastic resonance}.
\newblock \emph{Reviews of Modern Physics}, 70:\penalty0 223--287, January
  1998.
\newblock \doi{10.1103/RevModPhys.70.223}.

\bibitem[{Gosling} et~al.(2009){Gosling}, {McComas}, {Roberts}, and
  {Skoug}]{gosling09}
J.~T. {Gosling}, D.~J. {McComas}, D.~A. {Roberts}, and R.~M. {Skoug}.
\newblock {A One-Sided Aspect of Alfvenic Fluctuations in the Solar Wind}.
\newblock \emph{The Astrophysical Journal Letters}, 695:\penalty0 L213--L216,
  April 2009.
\newblock \doi{10.1088/0004-637X/695/2/L213}.

\bibitem[{Greco} et~al.(2009){Greco}, {Matthaeus}, {Servidio}, and
  {Dmitruk}]{greco09}
A.~{Greco}, W.~H. {Matthaeus}, S.~{Servidio}, and P.~{Dmitruk}.
\newblock {Waiting-time distributions of magnetic discontinuities: Clustering
  or Poisson process?}
\newblock \emph{Physical Review E}, 80\penalty0 (4):\penalty0 046401, October
  2009.
\newblock \doi{10.1103/PhysRevE.80.046401}.

\bibitem[{Horbury} et~al.(2018){Horbury}, {Matteini}, and {Stansby}]{horbury18}
T.~S. {Horbury}, L.~{Matteini}, and D.~{Stansby}.
\newblock {Short, large-amplitude speed enhancements in the near-Sunfast solar
  wind}.
\newblock \emph{Monthly Not. Royal Astron. Soc.}, 478:\penalty0 1980--1986,
  August 2018.
\newblock \doi{10.1093/mnras/sty953}.

\bibitem[{Horbury} et~al.(2020){Horbury}, Woolley, Laker, Matteini, Eastwood,
  Bale, Velli, Chandran, Phan, Raouafi, Goetz, Harvey, Pulupa, Klein, de~Wit,
  Kasper, Korreck, Case, Stevens, Whittlesey, Larson, MacDowall, Malaspina, and
  Livi]{horbury19}
Timothy~S. {Horbury}, Thomas Woolley, Ronan Laker, Lorenzo Matteini, Jonathan
  Eastwood, Stuart~D. Bale, Marco Velli, Benjamin D.~G. Chandran, Tai Phan,
  Nour~E. Raouafi, Keith Goetz, Peter~R. Harvey, Marc Pulupa, K.G. Klein,
  Thierry~Dudok de~Wit, Justin~C. Kasper, Kelly~E. Korreck, A.~W. Case,
  Michael~L. Stevens, Phyllis Whittlesey, Davin Larson, Robert~J. MacDowall,
  David~M. Malaspina, and Roberto Livi.
\newblock {Sharp Alfv\'enic impulses in the near-Sun solar wind}.
\newblock \emph{The Astrophysical Journal Supplement Series}, submitted, this
  volume, 2020.

\bibitem[Howard et~al.(2019)Howard, Vourlidas, Bothmer, Colaninno, DeForest,
  Gallagher, Hall, Hess, Higginson, Korendyke, Kouloumvakos, Lamy, Liewer,
  Linker, Linton, Penteado, Plunkett, Poirier, Raouafi, Rich, Rochus,
  Rouillard, Socker, Stenborg, Thernisien, and Viall]{howard19}
R.~A. Howard, A.~Vourlidas, V.~Bothmer, R.~C. Colaninno, C.~E. DeForest,
  B.~Gallagher, J.~R. Hall, P.~Hess, A.~K. Higginson, C.~M. Korendyke,
  A.~Kouloumvakos, P.~L. Lamy, P.~C. Liewer, J.~Linker, M.~Linton, P.~Penteado,
  S.~P. Plunkett, N.~Poirier, N.~E. Raouafi, N.~Rich, P.~Rochus, A.~P.
  Rouillard, D.~G. Socker, G.~Stenborg, A.~F. Thernisien, and N.~M. Viall.
\newblock {Near-Sun observations of an F-corona decrease and K-corona fine
  structure}.
\newblock \emph{Nature}, 2019.
\newblock \doi{10.1038/s41586-019-1807-x}.
\newblock URL \url{https://doi.org/10.1038/s41586-019-1807-x}.

\bibitem[Jagarlamudi et~al.(2019)Jagarlamudi, de~Wit, Krasnoselskikh, and
  Maksimovic]{jagarlamudi19}
Vamsee~Krishna Jagarlamudi, Thierry~Dudok de~Wit, Vladimir Krasnoselskikh, and
  Milan Maksimovic.
\newblock Inherentness of non-stationarity in solar wind.
\newblock \emph{The Astrophysical Journal}, 871\penalty0 (1):\penalty0 68, jan
  2019.
\newblock \doi{10.3847/1538-4357/aaef2e}.

\bibitem[{Jokipii} and {Parker}(1969)]{jokipii69}
J.~R. {Jokipii} and E.~N. {Parker}.
\newblock {Stochastic Aspects of Magnetic Lines of Force with Application to
  Cosmic-Ray Propagation}.
\newblock \emph{The Astrophysical Journal}, 155:\penalty0 777, March 1969.
\newblock \doi{10.1086/149909}.

\bibitem[{Kahler} et~al.(1996){Kahler}, {Crooker}, and {Gosling}]{kahler96}
S.~W. {Kahler}, N.~U. {Crooker}, and J.~T. {Gosling}.
\newblock {The topology of intrasector reversals of the interplanetary magnetic
  field}.
\newblock \emph{Journal of Geophysical Research}, 101:\penalty0 24373--24382,
  November 1996.
\newblock \doi{10.1029/96JA02232}.

\bibitem[{Kantelhardt} et~al.(2001){Kantelhardt}, {Koscielny-Bunde}, {Rego},
  {Havlin}, and {Bunde}]{kantelhardt01}
J.~W. {Kantelhardt}, E.~{Koscielny-Bunde}, H.~H.~A. {Rego}, S.~{Havlin}, and
  A.~{Bunde}.
\newblock {Detecting long-range correlations with detrended fluctuation
  analysis}.
\newblock \emph{Physica A Statistical Mechanics and its Applications},
  295:\penalty0 441--454, June 2001.
\newblock \doi{10.1016/S0378-4371(01)00144-3}.

\bibitem[Kasper et~al.(2019)Kasper, Bale, Belcher, Berthomier, Case, Chandran,
  Curtis, Gallagher, Gary, Golub, Halekas, Ho, Horbury, Hu, Huang, Klein,
  Korreck, Larson, Livi, Maruca, Lavraud, Louarn, Maksimovic, Martinovic,
  McGinnis, Pogorelov, Richardson, Skoug, Steinberg, Stevens, Szabo, Velli,
  Whittlesey, Wright, Zank, MacDowall, McComas, McNutt, Pulupa, Raouafi, and
  Schwadron]{kasper19}
J.~C. Kasper, S.~D. Bale, J.~W. Belcher, M.~Berthomier, A.~W. Case, B.~D.~G.
  Chandran, D.~W. Curtis, D.~Gallagher, S.~P. Gary, L.~Golub, J.~S. Halekas,
  G.~C. Ho, T.~S. Horbury, Q.~Hu, J.~Huang, K.~G. Klein, K.~E. Korreck, D.~E.
  Larson, R.~Livi, B.~Maruca, B.~Lavraud, P.~Louarn, M.~Maksimovic,
  M.~Martinovic, D.~McGinnis, N.~V. Pogorelov, J.~D. Richardson, R.~M. Skoug,
  J.~T. Steinberg, M.~L. Stevens, A.~Szabo, M.~Velli, P.~L. Whittlesey, K.~H.
  Wright, G.~P. Zank, R.~J. MacDowall, D.~J. McComas, R.~L. McNutt, M.~Pulupa,
  N.~E. Raouafi, and N.~A. Schwadron.
\newblock {Alfv{\'e}nic velocity spikes and rotational flows in the near-Sun
  solar wind}.
\newblock \emph{Nature}, 2019.
\newblock \doi{10.1038/s41586-019-1813-z}.
\newblock URL \url{https://doi.org/10.1038/s41586-019-1813-z}.

\bibitem[Koutsoyiannis(2003)]{koutsoyiannis03}
Demetris Koutsoyiannis.
\newblock Climate change, the hurst phenomenon, and hydrological statistics.
\newblock \emph{Hydrological Sciences Journal}, 48\penalty0 (1):\penalty0
  3--24, 2012/03/28 2003.
\newblock \doi{10.1623/hysj.48.1.3.43481}.

\bibitem[Krasnoselskikh et~al.(2020)Krasnoselskikh, Larosa, Agapitov, de~Wit,
  Moncuquet, Mozer, Stevens, Bale, Bonnell, Froment, Goetz, Goodrich, Harvey,
  Kasper, MacDowall, Malaspina, Pulupa, Raouafi, Revillet, Velli, , and
  Wygant]{krasnoselskikh19}
V.~Krasnoselskikh, A.~Larosa, O.~Agapitov, T.~Dudok de~Wit, M.~Moncuquet, F.~S.
  Mozer, M.~Stevens, S.~D. Bale, J.~Bonnell, C.~Froment, K.~Goetz, K.~Goodrich,
  P.~Harvey, J.~Kasper, R.~MacDowall, D.~Malaspina, M.~Pulupa, N.~Raouafi,
  C.~Revillet, M.~Velli, , and J.~Wygant.
\newblock {Localized magnetic field structures and their boundaries in the
  near-Sun solar wind from Parker Solar Probe measurements}.
\newblock \emph{The Astrophysical Journal Supplement Series}, submitted, this
  volume, 2020.

\bibitem[{Landi} et~al.(2006){Landi}, {Hellinger}, and {Velli}]{landi06}
S.~{Landi}, P.~{Hellinger}, and M.~{Velli}.
\newblock {Heliospheric magnetic field polarity inversions driven by radial
  velocity field structures}.
\newblock \emph{Geophysical Research Letters}, 33:\penalty0 L14101, July 2006.
\newblock \doi{10.1029/2006GL026308}.

\bibitem[{Lepreti} et~al.(2001){Lepreti}, {Carbone}, and {Veltri}]{lepreti01}
F.~{Lepreti}, V.~{Carbone}, and P.~{Veltri}.
\newblock {Solar Flare Waiting Time Distribution: Varying-Rate Poisson or
  L{\'e}vy Function?}
\newblock \emph{The Astrophysical Journal Letters}, 555:\penalty0 L133--L136,
  July 2001.
\newblock \doi{10.1086/323178}.

\bibitem[{Malaspina} et~al.(2020){Malaspina}, Halekas, Ber{\v c}i{\v c},
  Larson, Whittlesey, Bale, Bonnell, de~Wit, Ergun, Howes, Goetz, Goodrich,
  Harvey, MacDowall, Pulupa, Case, Kasper, Korreck, Livi, and
  Stevens]{malaspina19}
David~M. {Malaspina}, Jasper Halekas, Laura Ber{\v c}i{\v c}, Davin Larson,
  Phyllis Whittlesey, Stuart~D. Bale, John~W. Bonnell, Thierry~Dudok de~Wit,
  Robert~E. Ergun, Gregory Howes, Keith Goetz, Katherine Goodrich, Peter~R.
  Harvey, Robert~J. MacDowall, Marc Pulupa, Anthony~W. Case, Justin~C. Kasper,
  Kelly~E. Korreck, Roberto Livi, and Michael~L. Stevens.
\newblock {Plasma Waves near the Electron Cyclotron Frequency in the near-Sun
  Solar Wind}.
\newblock \emph{The Astrophysical Journal Supplement Series}, accepted, this
  volume, 2020.

\bibitem[{Matteini} et~al.(2014){Matteini}, {Horbury}, {Neugebauer}, and
  {Goldstein}]{matteini14}
L.~{Matteini}, T.~S. {Horbury}, M.~{Neugebauer}, and B.~E. {Goldstein}.
\newblock {Dependence of solar wind speed on the local magnetic field
  orientation: Role of Alfv{\'e}nic fluctuations}.
\newblock \emph{Geophysical Research Letters}, 41:\penalty0 259--265, January
  2014.
\newblock \doi{10.1002/2013GL058482}.

\bibitem[{Matteini} et~al.(2018){Matteini}, {Stansby}, {Horbury}, and
  {Chen}]{matteini18}
L.~{Matteini}, D.~{Stansby}, T.~S. {Horbury}, and C.~H.~K. {Chen}.
\newblock {On the 1/f Spectrum in the Solar Wind and Its Connection with
  Magnetic Compressibility}.
\newblock \emph{The Astrophysical Journal Letters}, 869:\penalty0 L32, December
  2018.
\newblock \doi{10.3847/2041-8213/aaf573}.

\bibitem[{Matteini} et~al.(2019){Matteini}, {Stansby}, {Horbury}, and
  {Chen}]{matteini19}
L.~{Matteini}, D.~{Stansby}, T.~S. {Horbury}, and C.~H.~K. {Chen}.
\newblock {The rotation angle distribution underlying magnetic field
  fluctuations in the 1/ f range of solar wind turbulent spectra}.
\newblock \emph{Nuovo Cimento C Geophysics Space Physics C}, 42\penalty0
  (1):\penalty0 16, Jan 2019.
\newblock \doi{10.1393/ncc/i2019-19016-y}.

\bibitem[McComas et~al.(2019)McComas, Christian, Cohen, Cummings, Davis, Desai,
  Giacalone, Hill, Joyce, Krimigis, Labrador, Leske, Malandraki, Matthaeus,
  McNutt, Mewaldt, Mitchell, Posner, Rankin, Roelof, Schwadron, Stone, Szalay,
  Wiedenbeck, Bale, Kasper, Case, Korreck, MacDowall, Pulupa, Stevens, and
  Rouillard]{mccomas19}
D.~J. McComas, E.~R. Christian, C.~M.~S. Cohen, A.~C. Cummings, A.~J. Davis,
  M.~I. Desai, J.~Giacalone, M.~E. Hill, C.~J. Joyce, S.~M. Krimigis, A.~W.
  Labrador, R.~A. Leske, O.~Malandraki, W.~H. Matthaeus, R.~L. McNutt, R.~A.
  Mewaldt, D.~G. Mitchell, A.~Posner, J.~S. Rankin, E.~C. Roelof, N.~A.
  Schwadron, E.~C. Stone, J.~R. Szalay, M.~E. Wiedenbeck, S.~D. Bale, J.~C.
  Kasper, A.~W. Case, K.~E. Korreck, R.~J. MacDowall, M.~Pulupa, M.~L. Stevens,
  and A.~P. Rouillard.
\newblock {Probing the energetic particle environment near the Sun}.
\newblock \emph{Nature}, 2019.
\newblock \doi{10.1038/s41586-019-1811-1}.
\newblock URL \url{https://doi.org/10.1038/s41586-019-1811-1}.

\bibitem[McManus et~al.(2020)McManus, Bowen, Mallet, Chen, Chandran, Bale, and
  Larson]{mcmanus19}
Michael~D. McManus, Trevor~A. Bowen, Alfred Mallet, Christopher H.~K. Chen,
  Benjamin D.~G. Chandran, Stuart~D. Bale, and Davin~E. Larson.
\newblock {Cross Helicity Reversals In Magnetic Switchbacks}.
\newblock \emph{The Astrophysical Journal Supplement Series}, submitted, this
  volume, 2020.

\bibitem[{Metzler} et~al.(2009){Metzler}, {Chechkin}, and {Klafter}]{metzler09}
Ralf {Metzler}, Aleksei~V. {Chechkin}, and Joseph {Klafter}.
\newblock {L{\'e}vy Statistics and Anomalous Transport: L{\'e}vy flights and
  Subdiffusion}.
\newblock In Robert~A. Meyers, editor, \emph{{Encyclopedia of Complexity and
  System Science}}, pages 1724--1745. Springer Verlag, New York, Jun 2009.
\newblock \doi{10.1007/978-0-387-30440-3}.

\bibitem[{Neugebauer}(2012)]{neugebauer12}
M.~{Neugebauer}.
\newblock {Evidence for Polar X-Ray Jets as Sources of Microstream Peaks in the
  Solar Wind}.
\newblock \emph{The Astrophysical Journal}, 750:\penalty0 50, May 2012.
\newblock \doi{10.1088/0004-637X/750/1/50}.

\bibitem[{Neugebauer} and {Goldstein}(2013)]{neugebauer13}
M.~{Neugebauer} and B.~E. {Goldstein}.
\newblock {Double-proton beams and magnetic switchbacks in the solar wind}.
\newblock In \emph{Proceedings of the Thirteenth International Solar Wind
  Conference}, volume 1539 of \emph{AIP Conference Proceedings}, pages 46--49,
  June 2013.
\newblock \doi{10.1063/1.4810986}.

\bibitem[Nicolis(1993)]{nicolis93}
C.~Nicolis.
\newblock Long-term climatic transitions and stochastic resonance.
\newblock \emph{Journal of Statistical Physics}, 70\penalty0 (1):\penalty0
  3--14, Jan 1993.
\newblock ISSN 1572-9613.
\newblock \doi{10.1007/BF01053950}.

\bibitem[{Parker}(1963)]{parker63}
E.~N. {Parker}.
\newblock \emph{{Interplanetary dynamical processes}}.
\newblock Interscience Publishers, New York, 1963.

\bibitem[{Passot} et~al.(2006){Passot}, {Ruban}, and {Sulem}]{passot06}
T.~{Passot}, V.~{Ruban}, and P.~L. {Sulem}.
\newblock {Fluid description of trains of stationary mirror structures in a
  magnetized plasma}.
\newblock \emph{Physics of Plasmas}, 13\penalty0 (10):\penalty0 102310, October
  2006.
\newblock \doi{10.1063/1.2356485}.

\bibitem[Peierls(1936)]{peierls36}
R.~Peierls.
\newblock {On Ising's model of ferromagnetism}.
\newblock \emph{Mathematical Proceedings of the Cambridge Philosophical
  Society}, 32\penalty0 (3):\penalty0 477--481, 1936.
\newblock \doi{10.1017/S0305004100019174}.

\bibitem[{Perrone} et~al.(2013){Perrone}, {Dendy}, {Furno}, {Sanchez},
  {Zimbardo}, {Bovet}, {Fasoli}, {Gustafson}, {Perri}, {Ricci}, and
  {Valentini}]{perrone13}
D.~{Perrone}, R.~O. {Dendy}, I.~{Furno}, R.~{Sanchez}, G.~{Zimbardo},
  A.~{Bovet}, A.~{Fasoli}, K.~{Gustafson}, S.~{Perri}, P.~{Ricci}, and
  F.~{Valentini}.
\newblock {Nonclassical Transport and Particle-Field Coupling: from Laboratory
  Plasmas to the Solar Wind}.
\newblock \emph{Space Science Reviews}, 178:\penalty0 233--270, October 2013.
\newblock \doi{10.1007/s11214-013-9966-9}.

\bibitem[{Pommois} et~al.(2001){Pommois}, {Veltri}, and {Zimbardo}]{pommois01}
P.~{Pommois}, P.~{Veltri}, and G.~{Zimbardo}.
\newblock {Field line diffusion in solar wind magnetic turbulence and energetic
  particle propagation across heliographic latitudes}.
\newblock \emph{Journal of Geophysical Research}, 106:\penalty0 24965--24978,
  November 2001.
\newblock \doi{10.1029/2001JA900050}.

\bibitem[Press et~al.(2002)Press, Teukolsky, Vetterling, and Flannery]{press02}
W.~H. Press, S.~A. Teukolsky, W.~T. Vetterling, and B.~P. Flannery.
\newblock \emph{Numerical Recipes: the Art of Scientific Computing}.
\newblock Cambridge University Press, Cambridge, 3rd edition, 2002.

\bibitem[Pulupa et~al.(2020)Pulupa, Bale, Badman, Bonnell, Case, de~Wit, Goetz,
  Harvey, Hegedus, Kasper, Korreck, Krasnoselskikh, Larson, Lecacheux, Livi,
  MacDowall, Maksimovic, Malaspina, Oliveros, Meyer-Vernet, Moncuquet, Stevens,
  and Whittlesey]{pulupa19}
Marc Pulupa, Stuart~D. Bale, Samuel~T. Badman, John~W. Bonnell, Anthony~W.
  Case, Thierry~Dudok de~Wit, Keith Goetz, Peter~R. Harvey, Alexander~M.
  Hegedus, Justin~C. Kasper, Kelly~E. Korreck, Vladimir Krasnoselskikh, Davin
  Larson, Alain Lecacheux, Roberto Livi, Robert~J. MacDowall, Milan Maksimovic,
  David~M. Malaspina, Juan Carlos~Martinez Oliveros, Nicole Meyer-Vernet,
  Michel Moncuquet, Michael Stevens, and Phyllis Whittlesey.
\newblock {Statistics and Polarization of Type III Radio Bursts Observed in the
  Inner Heliosphere}.
\newblock \emph{The Astrophysical Journal Supplement Series}, accepted, this
  volume, 2020.

\bibitem[{Sato} and {Klages}(2019)]{sato19}
Y.~{Sato} and R.~{Klages}.
\newblock {Anomalous Diffusion in Random Dynamical Systems}.
\newblock \emph{Physical Review Letters}, 122\penalty0 (17):\penalty0 174101,
  May 2019.
\newblock \doi{10.1103/PhysRevLett.122.174101}.

\bibitem[{Schreiber} and {Schmitz}(2000)]{schreiber00}
T.~{Schreiber} and A.~{Schmitz}.
\newblock {Surrogate time series}.
\newblock \emph{Physica D}, 142:\penalty0 346--382, August 2000.
\newblock \doi{10.1016/S0167-2789(00)00043-9}.

\bibitem[{Sorriso-Valvo} et~al.(2007){Sorriso-Valvo}, {Stefani}, {Carbone},
  {Nigro}, {Lepreti}, {Vecchio}, and {Veltri}]{sorriso-valvo07b}
L.~{Sorriso-Valvo}, F.~{Stefani}, V.~{Carbone}, G.~{Nigro}, F.~{Lepreti},
  A.~{Vecchio}, and P.~{Veltri}.
\newblock {A statistical analysis of polarity reversals of the geomagnetic
  field}.
\newblock \emph{Physics of the Earth and Planetary Interiors}, 164:\penalty0
  197--207, October 2007.
\newblock \doi{10.1016/j.pepi.2007.07.001}.

\bibitem[{Tenerani} and et~al.(2020)]{tenerani19}
A.~{Tenerani} and et~al.
\newblock {Magnetic field kinks and folds in the solar wind}.
\newblock \emph{The Astrophysical Journal Supplement Series}, submitted, this
  volume, 2020.

\bibitem[{van Ballegooijen} et~al.(1998){van Ballegooijen}, {Cartledge}, and
  {Priest}]{ballegooien98}
A.~A. {van Ballegooijen}, N.~P. {Cartledge}, and E.~R. {Priest}.
\newblock {Magnetic Flux Transport and the Formation of Filament Channels on
  the Sun}.
\newblock \emph{The Astrophysical Journal}, 501:\penalty0 866--881, July 1998.
\newblock \doi{10.1086/305823}.

\bibitem[{Velli} et~al.(2011){Velli}, {Lionello}, {Linker}, and
  {Miki{\'c}}]{velli11}
M.~{Velli}, R.~{Lionello}, J.~A. {Linker}, and Z.~{Miki{\'c}}.
\newblock {Coronal Plumes in the Fast Solar Wind}.
\newblock \emph{The Astrophysical Journal}, 736:\penalty0 32, July 2011.
\newblock \doi{10.1088/0004-637X/736/1/32}.

\bibitem[{Vio} et~al.(2013){Vio}, {Diaz-Trigo}, and {Andreani}]{vio13}
R.~{Vio}, M.~{Diaz-Trigo}, and P.~{Andreani}.
\newblock {Irregular time series in astronomy and the use of the Lomb-Scargle
  periodogram}.
\newblock \emph{Astronomy and Computing}, 1:\penalty0 5--16, February 2013.
\newblock \doi{10.1016/j.ascom.2012.12.001}.

\bibitem[{Wheatland} and {Litvinenko}(2002)]{wheatland02}
M.~S. {Wheatland} and Y.~E. {Litvinenko}.
\newblock {Understanding Solar Flare Waiting-Time Distributions}.
\newblock \emph{Solar Physics}, 211:\penalty0 255--274, December 2002.
\newblock \doi{10.1023/A:1022430308641}.

\bibitem[{Whittlesey} and et~al.(2020)]{whittlesey19}
P.~{Whittlesey} and et~al.
\newblock {Electron Strahl in switchbacks}.
\newblock \emph{The Astrophysical Journal Supplement Series}, submitted, this
  volume, 2020.

\bibitem[{Yamauchi} et~al.(2004){Yamauchi}, {Suess}, {Steinberg}, and
  {Sakurai}]{yamauchi04}
Y.~{Yamauchi}, S.~T. {Suess}, J.~T. {Steinberg}, and T.~{Sakurai}.
\newblock {Differential velocity between solar wind protons and alpha particles
  in pressure balance structures}.
\newblock \emph{Journal of Geophysical Research (Space Physics)}, 109:\penalty0
  A03104, March 2004.
\newblock \doi{10.1029/2003JA010274}.

\bibitem[{Zouganelis} et~al.(2004){Zouganelis}, {Maksimovic}, {Meyer-Vernet},
  {Lamy}, and {Issautier}]{zouganelis04}
I.~{Zouganelis}, M.~{Maksimovic}, N.~{Meyer-Vernet}, H.~{Lamy}, and
  K.~{Issautier}.
\newblock {A Transonic Collisionless Model of the Solar Wind}.
\newblock \emph{The Astrophysical Journal}, 606:\penalty0 542--554, May 2004.
\newblock \doi{10.1086/382866}.

\end{thebibliography}

\end{document}